\documentclass{JHEP3}

\usepackage{amsmath,amsfonts,amsthm,times,pst-node}

\usepackage{pstricks}

\reversemarginpar

\newcommand{\tsum}{{\textstyle\sum}}

\def \ignore#1 { {} } 

\def \Fig#1#2#3 {
\begin{figure}
\begin{center}
\scalebox{#2}{\includegraphics{#1.eps}}
\label{#1}
\end{center}
\caption{#3}
\end{figure}
}

\def \bea {\begin{eqnarray}}
\def \eea {\end{eqnarray}}
\def \bee {\begin{eqnarray*}}
\def \eee {\end{eqnarray*}}
\def \nn  {\nonumber}

\def \begm {\begin{multline}}
\def \endm  {\end{multline}}
\def \bega {\begin{align}}
\def \enda  {\end{align}}

\def \sp   { \ \ \ , \ \ \ }

\def \tsum  { {\textstyle \sum} }

\def \deg {{-\frac{1}{2b}}}
\def \tdeg {{-\tfrac{1}{2b}}}

\def \p {\partial}
\def \pp#1 {{\frac{\p}{\p #1}}}
\def \ppd#1 {{\frac{\p^2}{\p #1 ^2}} }

\def \g {\gamma}

\def \G {\Gamma}  
\def \e {\epsilon}
\def \a {\alpha}
\def \up {\Upsilon_b}

\def \rar {\rightarrow}

\def \bmu {\bar{\mu}}
\def \bm  {\bar{m}}

\def \bg {\bar{\gamma}}

\def \bz {\bar{z}}

\def \bx {\bar{x}}
\def \bp {\bar{\partial}}

\def \bJ {\bar{J}}

\def \Tr {{\rm Tr }\ }
\def \id {{\rm id } }
\def \sgn{ {\rm sgn} }

\def \Res {{\rm Res }}

\def \Z {\mathbb{Z}}

\def \R {\mathbb{R}}
\def \C {\mathbb{C}}

\def \la {\left\langle}
\def \ra {\right\rangle}

\def \asl {\widehat{s\ell_2}}
\def \SLC    {SL(2,\C)}
\def \SLR    {SL(2,\R)}
\def \USLR   {\widetilde{SL}(2,\R)}
\def \SLU    {SL(2,\R)/U(1)}
\def \H      {H_3^+}

\def \gsin#1#2 {\left[\!\! \begin{array}{c} #1 \\ #2 \end{array}\!\! \right]}

\def \Fsep { \ , \ }
\def \F32#1#2#3#4#5#6{{}_3F_2 \left(\left.\begin{array}{c}#1 \Fsep #2 \Fsep
    #3 \\ #4\Fsep #5
    \end{array} \right| #6 \right) }

\usepackage{graphics}\usepackage{epic}
\usepackage{pstricks,pst-coil,pst-node}

\def \mycoil#1#2{\pscoil[coilarm=0,linewidth=.05,coilaspect=0,coilwidth=.4,
coilheight=1.2](2.2,0) \rput[#1]{*0}(2.5,0){$#2$}}

\def \thecoil{\pscoil[coilarm=0,linewidth=.05,coilaspect=0,coilwidth=.4,
coilheight=1.2]}

\newlength{\floatlength}

\def \extline#1#2{\psline[linewidth=.1](2.2,0)\rput[#1]{*0}(2.5,0){$#2$}}

\def \intline#1#2#3{\pcline[linewidth=.1](0,0)(#3,0)
\pssetlength{\floatlength}{#3} \pssetlength{\floatlength}{.5\floatlength}
\rput[#1]{*0}(\floatlength,0.3){$#2$}}

\def\bdy#1{\pspolygon[linecolor=white,fillstyle=hlines,hatchwidth=.5pt,hatchsep=2pt]
(0,0)(#1,0)(#1,-.3)(0,-.3) 
\pcline(0,0)(#1,0)}

\def \degfield#1#2{\psdots[dotscale=1.5,dotstyle=*](0,0)
\rput[#1]{*0}(0.4,0.4){$#2$}}

\def \genfield#1#2{\psdots[dotscale=1.8,dotstyle=+,dotangle=45](0,0)
\rput[#1]{*0}(0.4,0.4){$#2$}}

\def \blockhorl#1#2#3#4#5#6#7{
\rput[l]{70}(2.5,0){\extline{l}{#3}}
\rput[l]{-70}(2.5,0){\extline{l}{#4}}
\rput[l]{110}(-2.5,0){\extline{r}{#1}}
\rput[l]{-110}(-2.5,0){\extline{r}{#2}}
\rput[l]{-90}(1.5,2.2){\mycoil{t}{\scriptstyle #6}}
\rput[l]{-90}(-1.5,2.2){\mycoil{t}{\scriptstyle #5}}
\rput[l]{180}(2.5,0){\intline{t}{#7}{5}}
}

\title{Solution of the $\H$ model on a disc}

\author{Kazuo Hosomichi
\\
Service de Physique Th\'eorique, CEA Saclay \\
91191 Gif-sur-Yvette Cedex\\
France \\
 {\tt Kazuo.Hosomichi@cea.fr }
}

\author{Sylvain Ribault
\\
Deutsches Elektronen-Synchrotron, Theory Division \\
Notkestrasse 85, Lab 2a \\
Hamburg 22603 \\
Germany  \\
 {\tt sylvain.ribault@desy.de }
}

\abstract{
We determine all the correlators of the $\H$ model on a disc with
$AdS_2$-brane boundary conditions in terms of correlators of Liouville
theory on a disc with FZZT-brane boundary conditions. We argue that
the Cardy-Lewellen constraints are weaker in the $\H$ model than in rational
conformal field theories due to extra singularities of the
correlators, but strong enough to uniquely determine the bulk
two-point function on a disc. We confirm our results by detailed
analyses of the bulk-boundary two-point function and of the boundary
two-point function. In particular we
find that, although the target space symmetry preserved by 
$AdS_2$-branes is the group $\SLR$, the open string states between two
distinct parallel
$AdS_2$-branes belong to representations of the universal covering group.
}

\preprint{
\hepth{0610117}\\
DESY 06-171 \\
SPhT-T06/109
 }

\makeatletter\let\default@color\current@color\makeatother 

\begin{document}


\section{Introduction and summary}


String theory in $AdS_3$ plays an important r\^ole in 
building string theory models of black
holes and cosmology, and in the $AdS$/CFT
correspondence. The $AdS_3$ space-time is interesting because it is Lorentzian,
non-compact, and curved; the theory is nevertheless expected to be
tractable thanks to the $\asl$ affine symmetry. However, the
Lorentzian feature is still a major technical hurdle. It can be
avoided by Wick rotation, which makes
space-time Euclidean while still non-compact and curved, and relates
the theory to the $\H$ model,
i.e. string theory in the Euclidean $AdS_3$, which still has the
$\asl$ symmetry.
Solving the $\H$ model is therefore a crucial technical step in the study of
string theory in $AdS_3$. By solving the model we mean determining its
spectrum and arbitrary correlators on arbitrary Riemann surfaces. 
As was shown in \cite{gaw89, gaw91}, the partition function and the correlators
can be defined, and in 
a few 
simple cases computed, 
within a path integral formulation.
The $\H$ model has also been studied using
the conformal bootstrap formalism \cite{bpz84}, which
exploits consistency constraints (like crossing symmetry) on these
correlators. These consistency constraints were shown to be
sufficient for fully determining the correlators of the $\H$ model on
a sphere, and explicitly computing the three-point function \cite{tes97a,tes01b}. 

The usefulness of the conformal bootstrap formalism for the $\H$ model
on a sphere was not obvious from the start, since the formalism was
developed for rational conformal field theories whereas the $\H$ model
is a non-rational CFT with a continuous spectrum.
However,  this formalism also brought about significant
progress in the case of Liouville theory on the sphere and on the disc.
Liouville theory is a simpler non-rational CFT which can be considered
as solved in the sense of the bootstrap formalism, because some
elementary correlators were explicitly computed, in terms of which all the
other correlators can in principle be deduced. 

Encouraged by these examples, one might expect the $\H$ model on a
disc to be solvable by means of the conformal bootstrap
formalism. As was first noticed in
\cite{pon02}, there is however a problem due to the presence of singularities in some
correlators. These singularities 
weaken the Cardy-Lewellen constraints \cite{cl91,lew92}, 
i.e. the conformal bootstrap equations on the disc. 
Such singularities are a consequence of the $\asl$ symmetry of
the model and are therefore also present in the $\H$ model on the
sphere, where they can however be circumvented by analytic continuation.

We will show how the $\H$ model on the disc can be solved in spite of
these singularities. The main tool which enables us to analyze the
singularities and solve the model is the $\H$-Liouville
relation, which was first established in the case of the sphere \cite{rt05}. 
In particular, our main result is a formula (\ref{main}) for arbitrary
correlators of the $\H$ model at level $k>2$ on the disc, in terms of
correlators of Liouville theory at parameter $b^2=(k-2)^{-1}$ on the
disc. Schematically, (\ref{main}) reads:
\bea
\la \prod_{a=1}^n \Phi^{j_a} \prod_{b=1}^m
{}_{r_{b-1,b}} \Psi^{\ell_{b}}{}_{r_{b,b+1}}\ra \propto
\la \prod_{a=1}^n V_{\a_a}
 \prod_{b=1}^m {}_{s_{b-1,b}}\left(B_{\beta_b}\right)_{s_{b,b+1}}
\prod_{a'=1}^{n'} V_\deg 
\prod_{b'=1}^{m'} B_\deg  \ra\ ,
\eea
where $\Phi^{j_a},\Psi^{\ell_b}$ are $\H$ bulk and boundary fields
with spins $j_a,\ell_b$ respectively, $V_{\a_a},B_{\beta_b}$
corresponding Liouville bulk and boundary fields with corresponding momenta
$\a_a,\beta_b$ respectively, and $V_\deg,B_\deg$ are extra degenerate
Liouville fields.
The boundary conditions are maximally symmetric in both theories, they
correspond to $AdS_2$ branes \cite{pst01} in $\H$ with parameters
$r$ and FZZT branes
\cite{fzz00,tes00} in Liouville theory with parameters
$s=\frac{r}{2\pi b} \pm \frac{i}{4b}$.
We are able to prove the
formula for all correlators which do not involve boundary condition
changing operators, leaving the remaining cases as a strongly
supported conjecture. We also reformulate our result suitably for its
application to the $\SLU$ coset model (\ref{mmain}). 

Since all Liouville correlators on the disc are known in principle,
our $\H$-Liouville relation on the disc amounts to a solution of the
$\H$ model on the disc. For some correlators, the conformal blocks are
simple enough that explicit expressions can be found. We will write such
explicit expressions in the cases of the boundary two-point function
(Section \ref{secbdy}) and the bulk-boundary two-point function
(Section \ref{secbulk}). These special cases will allow us to perform
some consistency checks: comparing the boundary two-point function
with predictions of the classical $\H$ model and with N=2 Liouville
theory, and the bulk-boundary two-point function with a minisuperspace
analysis. 

Finding the boundary two-point function 
amounts to determining the spectrum of open strings stretched between
two $AdS_2$ branes. In the case when these two branes are different,
we encounter a surprise: our results are incompatible with the $\SLR$
symmetry which was previously assumed for this system, and show that the
correct symmetry group is the universal covering group $\USLR$.

The Section \ref{secintro} and the Appendix provide supporting
material on the $\H$ model, special functions, and Liouville
theory. 

\section{The $\H$ model: state of the art \label{secintro}}

Let us review known results about the $\H$ model on Riemann surfaces
without boundaries, or with boundaries defined by $AdS_2$ branes. 

\subsection{Bulk $\H$ model}

The space $\H$ is a three-dimensional hyperboloid, or equivalently
the space of $(2\times2)$ Hermite matrices $h$ with unit determinant
and positive trace:
\bea
x_0^2-x_1^2-x_2^2-x_3^2 =1\ , \ ~
x_0>0\ ; \ ~~
h=\left(\begin{array}{cc} 
     x_0+x_3 & x_1-ix_2 \\  x_1+ix_2 & x_0-x_3 \end{array}\right)\ .
\eea
The $\H$ model at level $k$ on a two-dimensional Riemann surface $\Sigma$
parametrized by $z$ can be defined by the WZW-like action \cite{gaw91}
of a matrix field $h(z,\bar z)$, 
\bea
S^H[h]\,=\,
\frac{k}{2\pi}\int_\Sigma d^2z
          {\rm Tr}[h^{-1}\partial h h^{-1}\bar\partial h]
         +\frac{k}{12\pi i}\int_{\partial^{-1}\Sigma}
          {\rm Tr}(h^{-1}dh)^3 \ .
\eea
The $\H$ model is therefore a sigma-model with the manifold $\H$ as
target space, and a non-trivial $B$-field.
One often parametrizes $\H$ by coordinates $\phi,\gamma,\bar\gamma$ as
\bea
h=\left(\begin{array}{cc} 1&0\\ \gamma &1\end{array}\right)
 \left(\begin{array}{cc} e^\phi&0\\0 &e^{-\phi}\end{array}\right)
 \left(\begin{array}{cc} 1&\bar{\gamma}\\0 &1\end{array}\right)
=\left( \begin{array}{cc}e^\phi & e^\phi\bg \\ e^\phi\g & e^\phi
    \g\bg+e^{-\phi} \end{array}\right)\ .
\label{hmat}
\eea
In terms of these coordinates the action becomes
\begin{equation}
S^H\,=\,\frac k\pi \int d^2 z\; 
\left(\p \phi \bp \phi +e^{2\phi} \p \g \bp \bg \right)\ .
\label{H-act}
\end{equation}

The symmetry of the $\H$ model includes the $\SLC$ isometry group of the $\H$
manifold. The action of an $\SLC$ group element $g$ on $\H$ is
$g\cdot h\equiv ghg^\dagger$, so the element $g=-\id =
\left(\begin{array}{cc} -1 & 0 \\ 0 & -1 \end{array}\right)$ acts
trivially on $\H$. Thus the non-trivially acting isometry group is
actually $\SLC/\Z_2\simeq SO(1,3)$. The isometry group $SO(1,3)$
also follows  from the definition of $\H$ as an hyperboloid.

In the ``minisuperspace'' limit \cite{tes97b}, which involves sending
the level $k$ to infinity, the spectrum of the model reduces to
the space of functions on the $\H$ manifold
parametrized by $(\phi,\g,\bg)$. This
minisuperspace spectrum is generated by the following functions:
\bea
\Phi^{j}(x|h) =\frac{2j+1}{\pi}\left( [\begin{array}{cc} -x
    & 1 \end{array}] h \left[\begin{array}{c} -\bx \\ 1
  \end{array}\right]
 \right)^{2j}=
  \frac{2j+1}{\pi}\left(|\g-x|^2 e^\phi
+e^{-\phi}\right)^{2j} \ ,
\label{phijcl}
\eea
where delta-function normalizability requires $j\in
-\frac12+i\R$. This number $j$ is the spin of an $\SLC$
representation; states belonging to the same representation are
parametrized by the isospin variable $x\in \C$. The behaviour of
$\Phi^j$ under an $\SLC$ transformation $g=\left(\begin{array}{cc} \a
  &\g \\ \beta &\delta \end{array}\right)$ is
\bea
\Phi^j(x|g\cdot h) = |\gamma x-\delta|^{4j} \Phi^j(g\cdot
x|h) \ , \ g\cdot x =\frac{\a x-\beta}{\gamma x-\delta}\ .
\label{gx}
\eea
The spectrum of the quantum $\H$ model \cite{gaw91,tes97a} can
formally be built from the minisuperspace spectrum by acting with
oscillators encoding the worldsheet $z$-dependence, which amounts to
extending the representations of the group $\SLC$ into representations
of the corresponding loop group. The set of physical representations
itself does not change; unless specified otherwise our integrals on
the spin $j$ will be over this set $j\in -\frac12+i\R$.  
The conformal weight of the primary field $\Phi^j(x|z)$
built from the classical field $\Phi^{j}(x|h)$ is (using the
notation $b^2=(k-2)^{-1}$) 
\bea
\Delta_j=-b^2j(j+1)=-\frac{j(j+1)}{k-2}\ .
\eea
The symmetry algebra of the $\H$ model is (after
complexification) the affine Lie algebra $\asl\times \asl$ generated
by the modes of the currents $J=k\p hh^{-1}, \bJ=kh^{-1}\bp h$. 
This symmetry results in the correlators obeying the
Knizhnik--Zamolodchikov equations, which we will recall and
use in section \ref{secdisc}. For now, let us write the
consequences of the global symmetry group $\SLC$:
\bea
\la  \prod_{a=1}^n g\cdot \Phi^{j_a}(x_a|z_a) \ra = \la  \prod_{a=1}^n
 \Phi^{j_a}(x_a|z_a) \ra \ ,
\eea
where the $\SLC$ transformation of the quantum field is defined by
\bea
g\cdot \Phi^j(x|z) \equiv |\gamma x-\delta|^{4j} \Phi^j(g\cdot
x|z)\ .
\eea
Due to this simple transformation law, the isospin variable $x$ is
very convenient for the study of the 
$\SLC$ symmetry. 
But for the purpose of writing $\H$ correlators in terms
of Liouville correlators it is more convenient to use the
Fourier-transformed $\mu$-basis \cite{rt05}
\bea
\Phi^j(\mu|z) = \frac{1}{\pi}|\mu|^{2j+2}\int_\C d^2x\ e^{\mu x-\bmu \bx}
\Phi^j(x|z)\ .
\label{pmupx}
\eea
And for the purpose of comparing the $\H$ model with N=2
supersymmetric Liouville theory, we will need the $m$-basis
\bea
\Phi^j_{m\bm}(z) = \int \frac{d^2x}{|x|^2} x^{-j+m}\bx^{-j+\bm}\Phi^j(x|z) =
N^j_{m\bm} \int \frac{d^2\mu}{|\mu|^2} \mu^{-m}\bmu^{-\bm}
\Phi^j(\mu|z)\ ,
\label{phijmbm}
\eea
where the physical values of $m,\bm$ and the normalization
$N^j_{m\bm}$ are
\bea
m=\frac{n+ip}{2},\ \bm=\frac{-n+ip}{2},\ (n,p)\in \Z\times \R,\
N^j_{m\bm} =\frac{\G(-j+m)}{\G(j+1-\bm)}\ .
\label{njmbm}
\eea
Some basic correlators of the $\H$ model on a sphere can be written
explicitly. The bulk two-point function is
\begin{multline}
\la \Phi^{j_1}(\mu_1|z_1)\Phi^{j_2}(\mu_2|z_2) \ra =
|z_2-z_1|^{-4\Delta_{j_1}}  |\mu_1|^2\delta^{(2)}(\mu_2+\mu_1)  
\\
\times \big( 
\delta(j_2+j_1+1)+R^H(j_1)
\delta(j_2-j_1)\big)\ ,
\label{bulk2pt}
\end{multline}
where we introduce the bulk reflection coefficient $R^H(j)$ such that
\bea
\Phi^{j}(\mu|z)=R^H(j)\Phi^{-j-1}(\mu|z) \ , \
R^H(j)=b^{-2}\left(\frac{1}{\pi}b^2\gamma(b^2)\right)^{-(2j+1)} 
\frac{\gamma(+2j+1)}{\gamma(-b^2(2j+1))}\ .
\eea
The bulk three-point function  \cite{tes97a} is here written in the
$\mu$-basis in  a manifestly reflection-covariant way \cite{rib05a}:
\begin{multline}
\la\prod_{a=1}^3 \Phi^{j_a}(\mu_a|z_a)\ra
= \frac{ \delta^{(2)}(\tsum \mu_a)}{
|z_{12}|^{2\Delta^3_{12}}
|z_{13}|^{2\Delta_{13}^2}
|z_{23}|^{2\Delta^{1}_{23}}
}
D^H\bigg[\begin{array}{ccc} j_1 & j_2 & j_3 \\ \mu_1 & \mu_2
    &\mu_3  \end{array} \bigg]  C^H(j_1,j_2,j_3)\ , 
\\
D^H = \frac{|\mu_1|^{2j_1+2}}{|\mu_2|^{2j_1}}
\sum_{\eta=\pm}
\gamma_{j_3^\eta}^{j_1,j_2}\ 
\left|\frac{\mu_2}{\mu_3}\right|^{2j_3^\eta}\ {}_2{\cal F}
_1\big(j_1-j_2-j_3^\eta,j_1+j_2-j_3^\eta+1,-2j_3^\eta;-\frac{\mu_3}{\mu_2}\big)\
,
\\ \hspace{-28mm}
C^{H}= -\frac{1}{2\pi^2 b}
\left[\frac{\gamma(b^2)b^{2-2b^2}}{\pi}\right]^{-2-\Sigma j_i} 
\frac{\up'(0)}{\up(-b(j_{123}+1)) \G(-j_{123}-1)}\label{hb3pt}
\\
\times
\frac{\up(-b(2j_1+1))\up(-b(2j_2+1))\up(-b(2j_3+1))}
{\up(-bj_{12}^3) \G(-j_{12}^3)\ \up(-bj_{13}^2) \G(-j_{13}^2)\
\up(-bj_{23}^1)\G(-j_{23}^1)}\ .
\end{multline}
Notations:$
\left\{\begin{array}{l} 
\Delta^{3}_{12}=\Delta_{j_1}+\Delta_{j_2}-\Delta_{j_3},\ \
j_{12}^3=j_1+j_2-j_3,\ \ j_{123}=j_1+j_2+j_3,
\vspace{3mm}
\\
j^+=j,\ \
j^-=-j-1,\ \ {}_2{\cal F}_1(a,b,c;z)=F(a,b,c;z)F(a,b,c;\bar{z}),
\vspace{3mm}
\\
\gamma_{j_3}^{j_1,j_2}=
\G(-j_{123}-1) \G(-j_{23}^1) \G(-j_{13}^2) \G(j^3_{12}+1)
\gamma(2j_3+1).
\end{array}\right.$ \vspace{3mm} \linebreak
The special functions $\gamma$ and $\up$ are 
defined in the Appendix. The reflection
covariance of this expression follows from the reflection invariance
of $D^H$, and the reflection 
behaviour $C^H(j_1,j_2,j_3)=R^H(j_3)C^H(j_1,j_2,-j_3-1)$.

The four-point function of the $\H$ model has been shown to be
crossing symmetric \cite{tes01b}. This means that it can be deduced
from the three-point structure constant $C^H$ in two different ways:
\bea
\la \prod_{a=1}^4\Phi^{j_a}(\mu_a|z_a) \ra
&=& \int dj_s\ C^H(j_1,j_2,j_s)\ C^H(-j_s-1,j_3,j_4)\ {\cal G}^s_{j_s} (j_a|\mu_a|z_a)
\\
&=& \int dj_t\ C^H(j_1,j_4,j_t)\ C^H(-j_t-1,j_2,j_3)\ {\cal G}^t_{j_t}
(j_a|\mu_a|z_a)\ ,
\eea
where the $s$ and $t$-channel conformal blocks ${\cal G}^s_{j_s}
(j_a|\mu_a|z_a)$ and ${\cal G}^t_{j_t}
(j_a|\mu_a|z_a)$ are entirely determined by the affine $\asl$ symmetry
and thus in principle known before solving the model. This crossing
symmetry relation should be viewed as a constraint on the three-point
structure constant $C^H$. Exploiting very special cases of
this constraint was enough to unambiguously determine $C^H$
\cite{tes97a}. That this unique solution turned out to satisfy the full crossing
symmetry was an additional non-trivial check. 

\subsection{Euclidean $AdS_2$ branes}

Euclidean $AdS_2$ branes preserve an $\SLR$ subgroup of the bulk
symmetry group $\SLC$ \cite{pst01}.
 The geometry of these D-branes is
defined by the equation
\bea
\Tr \Omega h = 2\sinh r\ ,
\label{troh}
\eea
for $r$ a real parameter, and $\Omega$ a Hermitian matrix which
determines the relevant $\SLR$ subgroup as the set of $\SLC$ matrices such that 
$g^\dagger \Omega g =\Omega$. For definiteness we choose
$\Omega=\left(\begin{array}{cc} 0 & 1\\ 1 & 0 \end{array}\right)$, in
which case the $\SLR$ subgroup is the set of matrices
\bea
g=\left(\begin{array}{cc}  a & ic \\ -ib & d \end{array}\right),\
ad-bc=1,\ a,b,c,d\in \R\ .
\eea
In the minisuperspace limit, the spectrum of open strings on an
$AdS_2$ brane reduces to the space of functions on the corresponding
two-dimensional submanifold of $\H$. The minisuperspace spectrum is
generated by the functions:
\bea
\Psi^\ell(t|h)=\left( [\begin{array}{cc} it
    & 1 \end{array}] h \left[\begin{array}{c} -it \\ 1
  \end{array}\right]
 \right)^{\ell}\ ,
\label{psilcl}
\eea
where the boundary spin $\ell$ belongs to $-\frac12+i\R$, and
the boundary isospin is $t\in \R$. (For more details see Appendix A.2
of \cite{pst01}.) Under $\SLR$ transformations we have
\bea
\Psi^\ell(t|g\cdot h) = |ct-d|^{2\ell} \Psi^\ell(g\cdot t|h)\ ,\
g\cdot t=\frac{at-b}{-ct+d}\ .
\label{gpl}
\eea
The spectrum of the quantum model is generated by corresponding
boundary fields
$\Psi^\ell (t|w)$ with $w$ a real coordinate on the worldsheet
boundary, which transform as 
\bea
g\cdot \Psi^\ell (t|w) \equiv |ct-d|^{2\ell} \Psi^\ell(g\cdot t|w)\ .
\label{gplw}
\eea
There also exist $\SLR$ representations whose fields would behave as 
$ g\cdot \Psi^\ell (t|w) = |ct-d|^{2\ell} \sgn(-ct+d) \Psi^\ell(g\cdot t|w)$,
but such fields do not appear in the minisuperspace spectrum of
$AdS_2$ branes and we assume that they are absent from the exact
spectrum as well. We will naturally assume that correlators involving
boundary fields preserve the $\SLR $ symmetry:
\bea
\la  \prod_{a=1}^n g\cdot \Phi^{j_a}(x_a|z_a) \prod_{b=1}^m
g\cdot \Psi^{\ell_{b}}(t_{b}|w_{b})  \ra = \la  \prod_{a=1}^n
 \Phi^{j_a}(x_a|z_a)  \prod_{b=1}^m\Psi^{\ell_{b}}(t_{b}|w_{b}) \ra\
\ .
\label{slrsym}
\eea
We will also be interested in boundary condition changing fields ${}_{r'}
\Psi^\ell(t|w)_r $ describing open strings stretched between two
$AdS_2$ branes with different parameters $r,r'$. We will see in
Section \ref{secbdy} that the
symmetry properties of these fields are significantly more
complicated. So in the present review section we focus on the already
well-understood $r$-preserving fields.

The $t$-basis boundary fields we have considered so far are useful for
the study of the $\SLR$ symmetry. When it comes to the $\H$-Liouville
relation, it is more convenient to use the following $\nu$-basis
fields:
\bea
\Psi^\ell(\nu|w) =     |\nu|^{\ell+1} \int_\R dt\ e^{i\nu t}
\Psi^\ell(t|w)\ .
\label{pnupt}
\eea
The relation with the $\SLU$ coset and N=2 Liouville theory is more
naturally expressed using the $m$-basis fields, which diagonalize the
$t$-dilatations and $\nu$-dilatations: 
\bea
\Psi^\ell_{m,\eta}&=&\int_{-\infty}^\infty dt\
|t|^{-\ell-1+m}\sgn^\eta(t)\Psi^\ell(t) 
\\
&=& N^\ell_{m,\eta} \int \frac{d\nu}{|\nu|}
|\nu|^{-m}\sgn^\eta (\nu) \Psi^\ell(\nu) \ ,
\label{psiemeta}
\eea
where physical values of $m$ are pure imaginary, and we define
\bea
\eta\in\{0,1\} \sp N^\ell_{m,\eta} = 2i^\eta \G(-\ell+m)\sin
\tfrac{\pi}{2}(-\ell-1+m-\eta)  \ .
\label{nlmeta}
\eea
The boundary two-point function of open strings living on a single
$AdS_2$ brane of parameter $r$ is known to be
\cite{pst01}\footnote{Our formulas agree with \cite{pst01} only up to
  renormalization of the boundary fields.}
\begin{multline}
\la \Psi^{\ell_1}(t_1|w_1) \Psi^{\ell_2}(t_2|w_2)\ra_r
=|w_{12}|^{-2\Delta_{\ell_1}} 
\\ \times \frac{1}{2\pi} 
\left[ \delta(\ell_1+\ell_2+1)
\delta(t_{12}) + \delta(\ell_1-\ell_2) \tilde{R}_{r}^{H}(\ell_1)
|t_{12}|^{2\ell_1}  \right]
\label{ptpt}
\end{multline}
or equivalently 
\begin{multline}
\la \Psi^{\ell_1}(\nu_1|w_1) \Psi^{\ell_2}(\nu_2|w_2) \ra_r =
 |w_{12}|^{-2\Delta_{\ell_1}}
\\ \times  |\nu_1|\delta(\nu_1+\nu_2)
\left[\delta(\ell_1+\ell_2+1) + 
R^H_r(\ell_1)
  \delta(\ell_1-\ell_2) \right]\ .
\label{bdy2pt}
\end{multline}
The $\H$ boundary ``reflection number'' $\tilde{R} ^H_r(\ell)$ is related
to the $\H$ boundary reflection coefficient $R^H_r(\ell)$ by 
\bea
\tilde{R} ^H_r(\ell) = \frac{\pi}{\sin\pi \ell} \frac{1}{\G(2\ell+1)}
R^H_r(\ell) \ .
\eea
The quantity $R^H_r(\ell)$ deserves to be called the boundary
reflection coefficient because of its r\^ole in the simple reflection property of
the $\nu$-basis field,
\bea
\Psi^\ell(\nu|w)=R^H_r(\ell) \Psi^{-\ell-1}(\nu|w)\ .
\eea
Explicitly, $R^H_r(\ell)$ can be written in terms of the Liouville
boundary reflection coefficient (\ref{rlssp}), provided the Liouville
parameter is chosen as $b=(k-2)^{-1/2}$:
\bea
R^H_r(\ell)= R^L_{\frac{r}{2\pi b}-\frac{i}{4b},\frac{r}{2\pi
    b}+\frac{i}{4b}} \left(b(\ell+1)+\frac{1}{2b}\right) \ .
\label{rhrl}
\eea
This relation between the $\H$ and Liouville boundary reflection
coefficients is not surprising given the relation
$R^H(j)=R^L(b(j+1)+\tfrac{1}{2b})$ \cite{rt05} between bulk reflection
coefficients; the boundary relation actually follows from the relation
between the boundary states of the $AdS_2$ brane in $\H$ and the FZZT
brane in Liouville theory \cite{rib05b}, via the computation of the
annulus amplitude. 

Another known useful correlator is the bulk one-point function 
\cite{pst01,lpo01}
\begin{multline}
\la \Phi^j(x|z)\ra_r = \frac{1}{|z-\bz|^{2\Delta_j}}
\left[-\pi b^2\gamma(-b^2)\right]^{j+\frac12}
(8b^2)^{-\frac14 }
\\ \times 
|x+\bar{x}|^{2j}\Gamma(1+b^2(2j+1)) e^{-r(2j+1) \sgn
(x+\bar{x})} ,
\label{pxads}
\end{multline}
\vspace{-5mm}
\begin{multline}
\la \Phi^j(\mu|z)\ra_r=
\frac{1}{|z-\bar{z}|^{2\Delta_j}} \left[-\pi b^2\gamma(-b^2)\right]^{j+\frac12}
(8b^2)^{-\frac14 }
\\
\times |\mu|\delta(\Re \mu)  
\Gamma(2j+1)\Gamma(1+b^2(2j+1))  \cosh(2j+1)(r-i\tfrac{\pi}{2} \sgn \Im
\mu)\ .
\label{pmads}
\end{multline}


\section{$\H$ correlators on a disc \label{secdisc}}


Here we will study arbitrary $\H$ correlators on a disc.
We will express them  in terms of Liouville correlators,
which we consider as known quantities. The use of Liouville
correlators will become natural after we recall that the
Knizhnik--Zamolodchikov equations, which follow from the assumption 
that our $\H$ correlators preserve the affine Lie algebra symmetry of
the model, are equivalent to the Belavin--Polyakov--Zamolodchikov
equations satisfied by certain Liouville correlators. Due to the
existence of singularities, the KZ equations together with the usual
factorization axioms are not enough for fully determining the $\H$ correlators;
we will introduce the additional assumption of continuity at the
singularities. 
Then we will exhibit a
solution eq. (\ref{main}) of all these requirements in terms of Liouville
correlators. In the case of the bulk two-point function on the disc,
we will prove that this solution is unique, even though our continuity
assumption is weaker than the usual assumptions of the conformal
bootstrap formalism.

\subsection{Axioms for $\H$ correlators on a disc}

\subsubsection{Symmetry requirements}

We have already written the global $\SLR$ symmtry condition (\ref{slrsym}) 
for $\H$ correlators on a disc. Here we concentrate on the KZ
equations, which follow from the local $\asl$ symmetry. 
It was shown in \cite{rib05b} that the gluing
conditions for the $AdS_2$ branes are trivial in the $\mu$-basis,
which implies that the disc correlators satisfiy the same KZ equations
as the sphere correlators obtained by the ``doubling trick'', 
\bea
\la \prod_{a=1}^n \Phi^{j_a}(\mu_a|z_a) \prod_{b=1}^m
\Psi^{\ell_{b}}(\nu_{b}|w_{b})\ra_{\rm disc}  \rar 
\la \prod_{a=1}^n \left(\Phi^{j_a}(\mu_a|z_a)
\Phi^{j_a}(\bmu_a|\bz_a)\right) \prod_{b=1}^m \Phi^{\ell_b}(\nu_b|w_b)
\ra_{\rm sphere}\ .
\eea
The KZ equations for a bulk correlator $\Omega_n^H=\la \prod_{a=1}^n
\Phi^{j_a} (\mu_a|z_a)\ra $ are:
\begin{multline}
\left( (k-2)\pp{z_a} +\sum_{b\neq a} \frac{2t^3_a
  t^3_b-t^-_at^+_b-t^+_a t^-_b}{z_a-z_b} \right) \Omega^H_n =0\ ,
\left\{\begin{array}{l}
t^+_a=\mu_a \\ t^3_a=\mu_a\pp{\mu_a} \\ t^-_a=\mu_a\ppd{\mu_a}
  -\frac{j_a(j_a+1)}{\mu_a}\end{array}\right.\ .
\end{multline}
The power of these equations comes from the fact that they are first
order differential equations in $z_a$. So if we know a correlator at
some value of $z_1$ or in some limit say $z_1\rar z_2$, then the $\pp{z_1}
$ KZ equation determines that correlator for all values of $z_1$,
provided no singularities are met on the way. 

Explicit solutions of the KZ equations are known only in a few cases,
some of which we will see in Sections \ref{secbdy} and \ref{secbulk}. 
For our present purposes, it will however be enough to solve the KZ equations
in terms of Liouville correlators and conformal blocks. This is
possible thanks to the KZ-BPZ relation \cite{rt05,sto00}, which relates
the KZ equations for 
our $\H$ disc correlators to the BPZ equations satisfied
by certain Liouville disc correlators. 
We will denote this as a
relation $\simeq$ between $\H$ and Liouville disc correlators. (The KZ
and BPZ equations do not depend on the boundary conditions, which are
therefore omitted in the following formula.)
\begin{multline}
\la \prod_{a=1}^n \Phi^{j_a}(\mu_a|z_a) \prod_{b=1}^m
\Psi^{\ell_{b}}(\nu_{b}|w_{b})\ra 
\simeq  \delta\left(2\tsum_{a=1}^n \Re \mu_a +\tsum_{b=1}^m\nu_b\right)\ |u|
\ |\Theta_{n,m}|^\frac{k-2}{2} 
\\ \times
\la \prod_{a=1}^n V_{\a_a}
(z_a) \prod_{b=1}^m B_{\beta_b}(w_b) \prod_{a'=1}^{n'} V_\deg (y_{a'})
\prod_{b'=1}^{m'} B_\deg (y_{b'}) \ra\ ,
\label{kzbpz}
\end{multline}
where the $\H$ model at level $k$ is related to Liouville theory at
parameter $b$, background charge $Q$ and central charge $c_L$ with
\bea
b^2=\frac{1}{k-2} \ \ \ , \ \ \ Q=b+\frac{1}{b} \ \ \ , \ \ \ c_L=1+6Q^2
\ .
\label{bk}
\eea
The $\H$ spins $j,\ell$ are related to Liouville momenta $\a,\beta$ as
\bea
\a=b(j+1)+\frac{1}{2b}\ , \ \beta=b(\ell+1)+\frac{1}{2b}\ .
\label{ajbl}
\eea
The $n'$ bulk degenerate Liouville fields $V_\deg $ and $m'$ boundary
fields $B_\deg $ are introduced at positions determined by Sklyanin's
change of variables, which changes the isospin variables $\mu_a,\nu_b$
subject to the condition $2\sum_{a=1}^n \Re \mu_a +\sum_{b=1}^m
\nu_b=0$ (from global $s\ell(2)$ symmetry) into 
the variables $y_{a'},y_{b'}$ defined as the $2n'+m'=2n+m-2$ zeroes of the
function
\bea
\varphi(t) = \sum_{a=1}^n \frac{\mu_a}{t-z_a} +\sum_{a=1}^n
\frac{\bmu_a}{t-\bz_a} +\sum_{b=1}^m \frac{\nu_b}{t-w_b} \ ,
\label{vphi}
\eea
plus one real variable 
\bea
u=2\sum_{a=1}^n \Re ( \mu_a z_a) +\sum_{b=1}^m \nu_b w_b\ .
\label{udef}
\eea
The prefactor $\Theta_{n,m}$ is written in terms of
$Z_c=(z_a,\bz_a,w_b) $ and $Y_d=(y_{a'},\bar{y}_{a'},y_{b'})$ as
\bea
\Theta_{n,m} = \frac{\prod_{c<c'\leq 2n+m} (Z_c-Z_{c'})
  \prod_{d<d'\leq 2n+m-2} (Y_d-Y_{d'})
}{\prod_{c=1}^{2n+m}\prod_{d=1}^{2n+m-2} (Z_c-Y_d)}\ .
\label{thetadef}
\eea
We just provided enough data to make the relation (\ref{kzbpz}) 
between KZ and BPZ equations explicit. Let us give more details on
some relevant aspects and implications of this relation.

\paragraph{A closer look at Sklyanin's separation of variables.}

There is in general no explicit formula for the degenerate field positions $y$ as
functions of the isospin variables $\mu,\nu$. However, the definition
of $y$ as zeroes of
a function $\varphi(t)$ (\ref{vphi}) can be reformulated as
\bea
\varphi(t)=u \frac{\prod_{d=1}^{2n+m-2}
  (t-Y_d)}{\prod_{c=1}^{2n+m}(t-Z_c)}\ ,
\eea
which by
taking the limit $t\rar z_a$ or $t\rar w_b$ provides
an explicit formula for $\mu_a $ or $\nu_b$ in terms of $y$:
\bea
\mu_a &=& u \frac{\prod_{d=1}^{2n+m-2} (z_a-Y_d)}{(z_a-\bz_a)
  \prod_{a'\neq a\leq n} (z_a-z_{a'})(z_a-\bz_{a'}) \prod_{b=1}^m
  (z_a-w_b)} \ ,\nonumber
\\
\nu_b &=& u \frac{\prod_{d=1}^{2n+m-2}
  (w_b-Y_d)}{\prod_{a=1}^n|w_b-z_a|^2 \prod_{b'\neq b\leq m}
  (w_b-w_{b'})} \ .
\label{muofy}
\eea

\paragraph{Singularities of KZ solutions.} 

The KZ-BPZ relation (\ref{kzbpz}) allows us to easily study the singularities of the
KZ solutions, because the Liouville correlators on the right hand-side
are singular if and only if Liouville fields collide with each other
or with the boundary. If such a collision involves only the fields
$V_{\a_a}(z_a)$ and $B_{\beta_b}(w_b)$, then the corresponding
singularity at $z_a=z_{a'}, z_a=\bz_a$ or 
$w_b=w_{b'}$ is the power-like singularity expected from the $\H$ model
on general grounds. 

However, extra singularities occur where degenerate
Liouville fields $V_\deg(y_{a'})$ (or $B_\deg(y_{b'})$) are involved. 
If such a degenerate field comes close to  $V_{\alpha_a}(z_a)$ (or
$B_{\beta_b}(w_b)$), then $\varphi(t)$ loses its pole at $t=z_a$ which
implies $\mu_a=0$ (respectively, $\nu_b=0$). 
Such singularities will
play no significant r\^ole in the following, and should be considered
as artefacts of the $\mu$-basis. 
On the other hand, singularities
arising from collision of two boundary degenerate fields to become one
bulk degenerate field $B_\deg B_\deg \rar V_\deg $ (or vice versa)
will play a crucial
r\^ole\footnote{Such singularities are presumably equivalent to the
  ``$z=x$'' singularity in Fateev and Zamolodchikov's KZ-BPZ relation
\cite{fz86}  in the $x$-basis.}; in the following we will always refer
to these singularities when writing about singularities of $\H$
correlators. Let us explain their importance 
in the case of the bulk two-point function on the disc
$\la \Phi^{j_1}(\mu_1|z_1)\Phi^{j_2}(\mu_2|z_2)\ra$. (This case was
already studied in \cite{rib05b}.)

Given $\sum_{a=1}^2 \Re \mu_a=0$, the function
$\varphi(t)=\sum_{a=1}^2
\left(\frac{\mu_a}{t-z_a}+\frac{\bmu_a}{t-\bz_a} \right) $ has two
zeroes. If they are both real, they correspond to
two Liouville degenerate boundary fields in a correlator \linebreak
$\la V_{\a_1}(z_1)V_{\a_2}(z_2)B_\deg(y_1)B_\deg(y_2)\ra$: we call this situation the
{\it boundary regime}. If they are complex conjugate, they define the
position of one Liouville degenerate bulk field in a correlator $\la
V_{\a_1}(z_1)V_{\a_2}(z_2)V_\deg(y_1)\ra$: we call this the {\it
  bulk regime}. The positions of the Liouville fields involved in the
KZ-BPZ relation (\ref{kzbpz}) in the case of the $\H$ bulk two-point
function on the disc can be depicted as:
\bea
\begin{array}{c}
\psset{unit=.4cm}
\pspicture[](-3,-3.9)(5,2.9)
    \rput[l]{0}(-3.5,0){\bdy{7}}
    \rput[l]{-90}(-2.9,1.1){\genfield{l}{z_1}}
    \rput[l]{-90}(2.9,1.1){\genfield{l}{z_2}}
        \rput[l]{45}(-1,0){\degfield{l}{y_1}}
 \rput[l]{45}(1,0){\degfield{l}{y_2}}
       \rput[b](0,-3.5){Boundary regime}
\endpspicture
\pspicture[](-5,-3.9)(5,2.4)
    \rput[l]{0}(-3.5,0){\bdy{7}}
    \rput[l]{-90}(-2.5,2){\genfield{l}{z_1}}
    \rput[l]{-90}(2.5,2){\genfield{l}{z_2}}
        \rput[l]{-90}(0,0){\degfield{l}{}}
       \rput[b](0,-3.5){Singularity}
\endpspicture
\pspicture[](-5,-3.9)(8,2.4)
    \rput[l]{0}(-3.5,0){\bdy{7}}
    \rput[l]{-90}(-2,2.6){\genfield{l}{z_1}}
    \rput[l]{-90}(2,2.6){\genfield{l}{z_2}}
        \rput[l]{-90}(0,1){\degfield{l}{y_1}}
       \rput[b](0,-3.5){Bulk regime}
\endpspicture
\\
\psset{unit=.4cm}
\pspicture[](-13,-2.5)(18,0)
\psline[linewidth=1.6pt]{->}(-15,0)(15,0)
\rput[t](0,-.5){$\frac{|\mu_1|+|\mu_2|}{|\mu_1+\mu_2|}$}
\rput[t](-15,-.5){$1$}
\rput[t](14.5,-.5){$+\infty$}
\rput[l](16,0){$z=\left|\frac{z_1-\bz_2}{z_1-z_2}\right|$}
\psdots[dotstyle=|,dotscale=1.2](-15,0)(0,0)
\endpspicture
\end{array}
\label{bsb}
\eea
This singularity is significant because it separates two regimes which
are not otherwise connected, since the cross-ratio $z$ takes real
values. This is in contrast to the similar singularity which appears
in the $\H$ four-point function on a sphere. The related Liouville
correlator is in that case $\la\prod_{a=1}^4 V_{\a_a}(z_a)\ V_\deg(y_1)
V_\deg(y_2)\ra$, and one can go around the singularity $y_1=y_2$ by
moving $y_1,y_2$ in the Riemann sphere.

\subsubsection{Factorization axioms}

Factorization is a standard axiom of quantum field theory. It states
that in the limit where two of the fields come close, the correlator 
$\la \prod_{a=1}^n \Phi^{j_a}(\mu_a|z_a) \prod_{b=1}^m
\Psi^{\ell_{b}}(\nu_{b}|w_{b})\ra$ reduces to lower correlators
determined by the operator product expansion of the two fields. We of
course assume that $\H$ correlators obey such factorization
axioms. Note that factorization will only require taking limits of the
worldsheet positions $z_a,w_b$ of the fields, while their isospin
variables $\mu_a,\nu_b$ are kept fixed and arbitrary.

Depending on the nature of the two fields
which come close, there are three types of factorization, which
correspond to inserting the three types
of operator product expansions into the correlators : 
\begin{itemize}
\item Bulk OPE:
\begin{multline}
\Phi^{j_1}(\mu_1|z_1) \Phi^{j_2}(\mu_2|z_2) \underset{z_{12}\rar
  0}{\sim} \int dj\ \int \frac{d^2\mu}{|\mu|^2} 
|z-z_1|^{4\Delta_j}
\\
\la \Phi^{j_1}(\mu_1|z_1) \Phi^{j_2}(\mu_2|z_2)
\Phi^{-j-1}(-\mu|z) \ra
\times\left(\Phi^j(\mu|z_1) +O(z_{12})\right) \ ,
\label{bulkope}
\end{multline}
\item Bulk-boundary OPE:
\begin{multline}
\Phi^j(\mu|z) \underset{z-\bz\rar 0}{\sim} \int d\ell\
\int\frac{d\nu}{|\nu|} |w-z|^{2\Delta_\ell}
\\
\la \Phi^j(\mu|z) \Psi^{-\ell-1}(-\nu|w) \ra_r
\times \left({}_r\Psi^\ell(\nu|z)_r +O(z-\bz)\right) \ ,
\label{bbope}
\end{multline}
\item  Boundary OPE:
\begin{multline}
{}_{r_1}\Psi^{\ell_1}(\nu_1|w_1)_r\Psi^{\ell_2}(\nu_2|w_2)_{r_2} 
\underset{w_{12}\rar 0}{\sim} \int d\ell\ \frac{d\nu}{|\nu|} |w-w_1|^{2\Delta_\ell}
\\
\la _{r_1}\Psi^{\ell_1}(\nu_1|w_1)_r\Psi^{\ell_2}(\nu_2|w_2)_{r_2}
\Psi^{-\ell-1}(-\nu|w)_{r_1} \ra
\times\left( {}_{r_1}\Psi^\ell(\nu|w_1)_{r_2}+O(w_{12})\right)\ .
\label{bdyope}
\end{multline}
\end{itemize}
(Note that the OPEs do not depend on the choice of the auxiliary
worldsheet variables $z,w$.)

We can formally write these OPEs without knowing the three basic correlators
(bulk three-point, bulk-boundary two-point, boundary three-point
functions); on the other hand we rely on the previous knowledge of the
bulk and boundary spectra and two-point 
functions\footnote{Although we do not yet know the spectrum of boundary fields
  ${}_{r_1}\Psi^\ell(\nu|w)_{r_2}$ when $r_1\neq r_2$, we assume that
  such fields are parametrized by the same values of $\ell$ and $\nu$
  as in the case $r_1=r_2$, and do not have additional indices.}
eq. (\ref{bulk2pt},\ref{bdy2pt}). 

Once inserted into a correlator, such
an OPE should be considered as a formal limit, since the corrections
$O(z_{12})$ to one term $j$ can be dominant with respect to the
leading contribution $\Phi^{j'}(\mu|z_1)$ of another term $j'$ of
higher conformal dimension. This formal limit consists in focussing on
the contribution of primary fields, and the corrections correspond to
descendants. Such corrections are in principle determined by the
symmetry of the model, in our case the affine Lie algebra symmetry. 

\paragraph{Factorization and Cardy-Lewellen formalism.}

We now discuss the crucial issue of the strength of the factorization
constraints, i.e. in which measure they determine the
correlators. First note that
if the sum over all descendant contributions 
converged for any values of the worldsheet variables, 
then the correlators would be fully determined by their
behaviour in one given factorization limit. For example, we would fully know
the bulk two-point 
function on the disc thanks to the limit where it reduces to the known
bulk three-point function on a sphere and bulk one-point function on
the disc:
\begin{multline}
\la \Phi^{j_1}(\mu_1|z_1)\Phi^{j_2}(\mu_2|z_2)\ra_r \underset{z_{12}\rar
  0}{\sim} \int dj\ \int \frac{d^2\mu}{|\mu|^2} 
|z-z_1|^{4\Delta_j}
\\
\la \Phi^{j_1}(\mu_1|z_1) \Phi^{j_2}(\mu_2|z_2)
\Phi^{-j-1}(-\mu|z) \ra 
\times\left(\la \Phi^j(\mu|z_1)\ra_r +O(z_{12})\right)
\end{multline}
We could now study $\la
\Phi^{j_1}(\mu_1|z_1)\Phi^{j_2}(\mu_2|z_2)\ra_r$ in the limit $z_1\rar
\bz_1$. Whether it would factorize or not would be a consistency test
on the bulk three-point and disc one-point functions. If the test was
passed, we could then deduce the bulk-boundary two-point function.
Such constraints and relations for structure constants were
systematically studied by Cardy and Lewellen \cite{cl91,lew92}.

The Cardy-Lewellen formalism actually applies in the cases of Liouville
theory and of the $\H$ model on the sphere.
In the latter case, 
the sums of descendant contributions however do not converge for all
values of the worldsheet variables $z$ (as is apparent from the
existence of singularities), but only in neighbourhoods of the various
factorization limits. But 
the affine Lie algebra symmetry which in principle determines
these sums actually yields a more powerful tool: the KZ equations. These
equations can  be used to analytically continue the correlators in regions
where the sums of descendants do not converge.

On the disc however,
the $\H$ bulk two-point function is not fully
determined by its behaviour near $z_{12}\rar 0$, because as shown in 
the picture (\ref{bsb}) it is impossible to go around the singularity.
We would need as additional data the
behaviour near $z_1-\bz_1\rar 0$, and therefore the (as yet unknown)
bulk-boundary two-point function\footnote{The situation is even worse in the
case of the boundary four-point function
$\la \prod_{b=1}^4 \Psi^{\ell_b}(\nu_b|w_b) \ra$: even if we knew the boundary
three-point function and therefore the behaviour in both possible
factorization limits $w_{12}\rar 0, w_{23}\rar 0$, we could not deduce
the boundary four-point function in the regime where the corresponding
Liouville correlator has one bulk degenerate field.}.
In terms of sums of descendants, the situation is presumably the following: the
sum of descendants in the bulk-boundary OPE converges near
$z_1=\bz_1$ and in the vicinity (up to the singularity), 
and therefore in the boundary
regime. The sum of descendants in the bulk OPE converges near
$z_1=z_2$ and in the vicinity (up to the singularity), 
and therefore in the bulk regime. But
the strength of the Cardy-Lewellen constraints relies on the
existence of an overlap between the domains of convergences of these
two OPEs. Such an overlap is absent in our case, as opposed to the
case of the bulk two-point function on the disc in Liouville
theory, where the sums of descendants in both OPEs converge for any
values of $z_1,z_2$ as was established in
\cite{tes03b} (Section 2.4 therein):
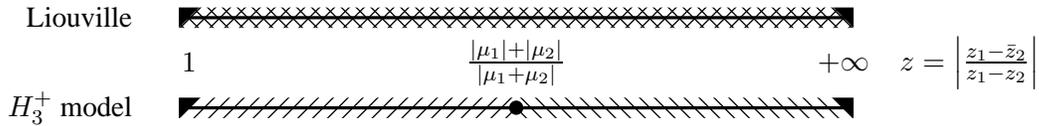
\bee
\psset{unit=.3cm}
\pspicture[](-18,-3)(18,5)
\psline[linewidth=1.2pt](-15,0)(15,0)
\psline[linewidth=1.2pt](-15,4)(15,4)
\rput[r](-17,0){$\H$ model}
\rput[r](-17,4){Liouville}
\rput(0,2){$\frac{|\mu_1|+|\mu_2|}{|\mu_1+\mu_2|}$}
\rput(-14.5,2){$1$}
\rput(14.5,2){$+\infty$}
\rput[l](17,2){$z=\left|\frac{z_1-\bz_2}{z_1-z_2}\right|$}
\pspolygon*(-15,-.5)(-15,.5)(-14,.5)
\pspolygon*(-15,3.5)(-15,4.5)(-14,4.5)
\pspolygon*(15,-.5)(15,.5)(14,.5)
\pspolygon*(15,3.5)(15,4.5)(14,4.5)
\pspolygon[fillstyle=crosshatch,hatchwidth=.5pt,hatchsep=3.5pt,linecolor=white]
(-15,3.5)(-15,4.5)(15,4.5)(15,3.5)
\pspolygon[fillstyle=hlines,hatchwidth=.5pt,hatchsep=3.5pt,linecolor=white]
(-15,-.5)(-15,.5)(0,.5)(0,-.5)
\pspolygon[fillstyle=vlines,hatchwidth=.5pt,hatchsep=3.5pt,linecolor=white]
(15,-.5)(15,.5)(0,.5)(0,-.5)
\psdots[dotscale=1.5,dotstyle=*](0,0)
\endpspicture
\eee
(Here the triangles denote the factorization limits, and the hatches
the corresponding regions where the sums of descendants converge.)

In this sense, the Cardy-Lewellen formalism does not fully apply to the $\H$ model
on the disc because of the singularities of the $\H$
correlators. Nevertheless, we can recover part of the power of the
Cardy-Lewellen constraints by making a natural assumption on the behaviour
of the $\H$ correlators at the singularities.

\subsubsection{Continuity assumption}

In contrast to the symmetry requirements and factorization axioms,
which are standard assumptions of conformal field theory in the
conformal bootstrap formalism, our continuity assumption will be a
novelty of the $\H$ model on the disc. Such an
assumption is made necessary by the existence of extra singularities
of the model (\ref{bsb}): for the formalism to be of any use, we need some control
over the behaviour of correlators at these singularities.

{\it
\paragraph{Continuity assumption:}
The $\H$ correlators are continuous at the
singularities which occur when degenerate fields in the corresponding
Liouville correlators collide.
}
\vspace{3mm}

\noindent In order to clarify the meaning of this assumption, let us recall
how KZ solutions behave near such singularities. This can easily be
deduced from the relevant Liouville OPEs, dressed with the 
$|y_{12}|^{\frac{k-2}{2}}$
prefactor from the KZ-BPZ relation (\ref{kzbpz}),
\bea
|y_{12}|^{\frac{k-2}{2}} B_\deg(y_1) B_\deg(y_2) &\underset{y_{12}\rar
  0}{\sim}&B_{-\frac{1}{b}}(y_1)+
C^L(\deg,\deg,Q) |y_{12}|^{2k-3} B_0(y_1) \ ,
\label{degbdyope}
\\
|y_{12}|^{\frac{k-2}{2}} V_\deg(y_1) &\underset{y_1-\bar{y}_1\rar
  0}{\sim}& B_{-\frac{1}{b}}(y_1) + B^L(\deg,Q)|y_{12}|^{2k-3} 
B_0(y_1) \ ,
\label{degbbope}
\eea
where we omit the dependences on the boundary parameters of the
Liouville boundary three-point function $C^L(\deg,\deg,Q)$,
bulk-boundary two-point function $B^L(\deg,Q)$, and boundary
fields. (Explicit formulas for the relevant OPE coefficients can be
found in the Appendix, eq. (\ref{bbbss}) and (\ref{vbbbb}).)

The leading behaviour of the KZ solutions therefore consists of two
terms, associated with the Liouville boundary fields $B_0$ and
$B_{-\frac{1}{b}}$. (The corrections to the leading behaviour are due
to descendants of these two fields.) The critical exponent of the
$B_{-\frac{1}{b}}$ term is zero, so such a term has a
finite limit whether it arises from the bulk regime ($V_\deg$ case) or
from the boundary regime ($B_\deg B_\deg$ case). The critical exponent
of the $B_0$ term is $2k-3>1$, such a term goes to zero at
the singularity. Therefore, all KZ solutions have finite limits at the
singularity. Our continuity assumption means that the limit
evaluated from the bulk regime should agree with the limit evaluated
from the boundary regime. This seems to us a very natural assumption.

Thus, the continuity assumption will be a
nontrivial requirement on $\H$ correlators, although it of course
does not fully determine how KZ solutions behave through the
singularity, because the $B_0$ term  remains
unconstrained. 

\subsection{$\H$ disc correlators from Liouville theory}

It is relatively easy to find an Ansatz for the $\H$ disc correlators
which satisfies all our axioms. The difficulty will be to prove that
the solution is unique. Let us first write our Ansatz for arbitrary
$\H$ correlators on the disc:
\begin{multline}
\la \prod_{a=1}^n \Phi^{j_a}(\mu_a|z_a) \prod_{b=1}^m
{}_{r_{b-1,b}}\Psi^{\ell_{b}}(\nu_{b}|w_{b})_{r_{b,b+1}}\ra 
\\ \hspace{-3.5cm}
= \pi^2\sqrt{\tfrac{b}{2}} (-\pi)^{-n}\ \delta\left(2\tsum_{a=1}^n \Re \mu_a +
\tsum_{b=1}^m\nu_b\right)\ |u|\
|\Theta_{n,m}|^\frac{k-2}{2} 
\\ \times
\la \prod_{a=1}^n V_{\a_a}
(z_a) \prod_{b=1}^m {}_{s_{b-1,b}}B_{\beta_b}(w_b)_{s_{b,b+1}}
\prod_{a'=1}^{n'} V_\deg (y_{a'}) 
\prod_{b'=1}^{m'} B_\deg (y_{b'}) \ra\ ,
\label{main}
\end{multline}
where most notations were already defined in our study of the KZ-BPZ
relation: the Liouville parameter $b$ (\ref{bk}), the Liouville
momenta $\alpha,\beta$ (\ref{ajbl}), the quantity $u$ (\ref{udef}),
the prefactor $\Theta_{n,m}$ (\ref{thetadef}). The positions
$y_{a'},y_{b'}$ of the
Liouville degenerate fields were defined as the zeroes of a function
$\varphi(t)$ (\ref{vphi}). In addition, we specify the set of
boundary conditions $s_{b-1,b}$ by 
\bea
s = \frac{r}{2\pi b} - \frac{i}{4b} \sgn \varphi(t) \ .
\label{srphi}
\eea
That is, the Liouville boundary parameter $s$ on a point $t$ of the
boundary is given by the $\H$ boundary parameter $r$, shifted by a
quantity which depends on $\sgn \varphi(t)$. (Indeed $\varphi(t)$ is
real if $t$ is real.) Notice that $\varphi(t)$ changes sign at its
zeroes, which are the positions of the boundary degenerate fields, and
when it is infinite, which happens at the points where the generic
boundary fields ${}_{s_{b-1,b}}B_{\beta_b}(w_b)_{s_{b,b+1}}$ are
inserted. 
So each boundary degenerate field $B_\deg (y_{b'})$ induces a jump
$\pm \tfrac{i}{2b}$ of the boundary parameter $s$, consistently with
the results
of Fateev, Zamolodchikov and Zamolodchikov \cite{fzz00}. 
Then, for a
given $AdS_2$ brane parameter $r$, there correspond two opposite values of the
Liouville boundary cosmological constant,
\bea
\mu_B=\sqrt{\frac{\mu_L}{\sin \pi b^2}} \cosh 2\pi b s= \pm
\sqrt{\frac{\mu_L}{\sin \pi b^2}} \sinh r\ .
\label{mub}
\eea
The formula (\ref{main}) is our main result and the rest of the
article is devoted to giving evidence for it, and drawing some
consequences. 

The first check is the compatibility with  the bulk
one-point function, which is explicitly known (\ref{pmads}). 
This check is straighforward and 
was already performed in \cite{rib05b}.

Let us check that our formula satisfies the axioms of the $\H$
model. By construction, our Ansatz (\ref{main}) satisfies the KZ
equations.
It is continuous at the singularities due to the agreement between the
coefficients of the leading terms of the Liouville
$\underset{y_{12}\rar 0}{\lim}B_\deg(y_1) B_\deg(y_2)$ and
$\underset{z_1-\bz_1\rar 0}{\lim} V_\deg(z_1)$ OPEs
(\ref{degbdyope}), (\ref{degbbope}). 
The only subtle issues come from the
factorization axioms:
\begin{itemize}
\item 
Bulk factorization $z_{12} \rar 0$: the pole $t=z_1$ of the function
$\varphi(t)$ (\ref{vphi}) must remain simple, so that
one Liouville bulk
  degenerate field say $V_\deg(y_1)$ must come close to
  $V_{\a_1}(z_1)$ and $V_{\a_2}(z_2)$, i.e. $y_1-z_1\propto z_{12}\rar 0$. 
Thus, we should insert into our Ansatz
  (\ref{main}) the following Liouville OPE:
\begin{multline}
V_{\a_1}(z_1)V_{\a_2}(z_2)V_\deg (y_1) \underset{z_{12}\propto z_1-y_1
  \rar 0}{\sim} \int d\a\ |z-z_1|^{4\Delta_\a}
\\
\la V_{\a_1}(z_1)V_{\a_2}(z_2)V_\deg (y_1)
  V_{Q-\a}(z) \ra 
\times
\left(V_\a(z_1) + {\cal O}(z_{12}) \right) \ ,
\end{multline}
where $\Delta_\a=\a(Q-\a)$ is the conformal dimension of a
Liouville field of momentum $\a$.
This is the crucial step in proving that our Ansatz indeed satisfies
the bulk OPE axiom (\ref{bulkope}), as was shown in detail in
\cite{rt05} in the case of $\H$ correlators on the sphere.
\item
Bulk-boundary factorization $z_1-\bz_1\rar 0$: by a similar reasoning,
one Liouville boundary
degenerate field say $B_\deg(y_1)$ must come close to
$V_{\a_1}(z_1)$. We should insert into our Ansatz
  (\ref{main}) the following Liouville OPE:
\begin{multline}
V_{\a_1}(z_1)B_\deg(y_1) \underset{z_1-\bz_1\propto z_1-y_1 \rar
  0}{\sim} \int d\beta\ |w-z_1|^{2\Delta_\beta} 
\\
\la V_{\a_1}(z_1)
  B_\deg(y_1) {}_{s_-} B_{Q-\beta}(w)_{s_+} \ra 
\times \left({}_{s_-}B_\beta(z_1)_{s_+}+{\cal O}(z_1-\bz_1)\right) \ ,
\end{multline}
with $s_\pm = \frac{r}{2\pi
  b}\pm\frac{i}{4b}\sgn(\mu_1+\bmu_1) $, and $r$ is the $\H$ boundary
  parameter at the point where $z_1$ reaches the boundary. Then one
  can check that the Liouville correlator \linebreak $ \la V_{\a_1}(z_1)
  B_\deg(y_1) {}_{s_-} B_{Q-\beta}(z_1)_{s_+} \ra$ agrees with the
  prediction of our Ansatz (\ref{main}) for the $\H$ bulk-boundary
  two-point function appearing in the $\H$ bulk-boundary OPE
  (\ref{bbope}).
\item
Boundary factorization $w_{12}\rar 0$: by a similar reasoning, 
one Liouville boundary
degenerate field say $B_\deg(y_1)$ must come close to
$B_{\beta_1}(w_1),B_{\beta_2}(w_2)$. We should insert into our Ansatz
  (\ref{main}) the following Liouville OPE:
\begin{multline}
B_{\beta_1}(w_1) B_{\beta_2}(w_2)B_\deg(y_1) \underset{w_{12}\propto y_1-w_1\rar
  0}{\sim} \int d\beta\ |w-w_1|^{2\Delta_\beta} 
\\
\la B_{\beta_1}(w_1)  B_{\beta_2}(w_2) B_\deg(y_1)  B_{Q-\beta}(w)\ra 
\times \left(B_\beta(w_1) +{\cal O}(w_{12})\right) \ ,
\end{multline}
where for definiteness we assumed the degenerate field to come on the
right on $B_{\beta_1}(w_1)$ and $B_{\beta_2}(w_2)$, while it may also
come on the left or in between, depending on the signs of
$\nu_1,\nu_2$ and $\nu_1+\nu_2$. For simplicity,
we omit the Liouville boundary parameters, which can easily be deduced
from our Ansatz.
This is the main step in checking that our Ansatz (\ref{main}) is compatible with
the $\H$ boundary OPE (\ref{bdyope}). 
\end{itemize}
There is however a property which we have not checked: the $\SLR$
group symmetry (\ref{slrsym}), or equivalently its Lie algebra version
$s\ell(2,\R)$. In the absence of boundaries, this symmetry is necessary for the
KZ-BPZ relation \cite{rib05a,rt05}, and is therefore
automatically included in the $\H$-Liouville relation. However, it is
not obvious that our Ansatz is $s\ell(2,\R)$ symmetric, because the
Liouville boundary parameter (\ref{srphi}) varies along the boundary,
in a way which is non-trivially affected by $s\ell(2,\R)$
transformations.
In the case of the bulk-boundary two-point
function (Section \ref{secbulk}), we will explicitly 
check the $\SLR$ symmetry of our Ansatz.

\subsection{Uniqueness of the solution to the axioms }

We have easily checked that our formula (\ref{main}) for the $\H$ disc
correlators verifies our axioms of symmetry, factorization and
continuity. We will now argue that this solution is unique in
the particular case of correlators with no boundary condition changing operators.
 
We will write an explicit argument only in the case of the bulk
two-point function on the disc. This will be enough to address the
crucial issue of the singularity separating the bulk and boundary
regimes, as defined in (\ref{bsb}). Let us spell out the formula to be
proved:
\begin{multline}
\la \Phi^{j_1}(\mu_1|z_1)\Phi^{j_2}(\mu_2|z_2)\ra_r = \sqrt{\tfrac{b}{8}}
\delta \left(\Re (\mu_1+\mu_2)\right)\
|u|\ \left( \frac{|z_{12}|^2|y_{12}| \prod_a |z_a-\bz_a|^2  }{
  \prod_{a,b}|z_a-y_b|^2} \right)^{\frac{k-2}{2}}
\\
\times \left\{ \begin{array}{ll} \la
  V_{\a_1}(z_1)V_{\a_2}(z_2)V_\deg(y_1)\ra_{s_+}  & {\rm if}\ y_2=\bar{y}_1\
  {\rm (bulk\ regime)}\ ,
\\ \la  V_{\a_1}(z_1)V_{\a_2}(z_2){}_{s_+}
B_\deg(y_1)_{s_-}B_\deg(y_2)_{s_+} \ra & {\rm
  if}\ y_1<y_2\in \R\ {\rm (boundary\ regime)}\ , \end{array} \right. 
\label{twoptcase}
\end{multline}
where $s_\pm=\frac{r}{2\pi b} \mp\frac{i}{4b} \sgn u $ with
$u=2\Re(\mu_1z_1+\mu_2z_2)$, and in the bulk regime we have \linebreak
$\sgn u = \sgn \Im (\mu_1+\mu_2)$.

The explicit knowledge of the $\H$ bulk one-point function on the
disc, and the axiom of bulk factorization (\ref{bulkope}), are enough
to prove the formula (\ref{twoptcase}) in the limit $z_{12}\rar
0$. Then, the local $\asl$ symmetry requirement and the knowledge that
the resulting KZ equations are equivalent to BPZ equations
(\ref{kzbpz}) show that the formula is true in the whole bulk
regime. 

The continuity assumption will now provide some information on the
bulk two-point function at the $z=\frac{|\mu_1|+|\mu_2|}{|\mu_1+\mu_2|}$
end of the boundary regime. The other end $z=1$ is constrained by the
axiom of bulk-boundary factorization (\ref{bbope}), which is a non-trivial
requirement even though we do not know the bulk-boundary two-point
function. These two limiting regions are connected by the KZ
equations, which hold in the whole boundary regime. We purport to show
that, taken toghether, these constraints are enough to fully determine
the bulk two-point function in the boundary regime.

The reasoning could now go in two possible directions, depending on
which one of the two limiting regions we consider first. If we first solve the
continuity assumption, it is then difficult to exploit the axiom of
bulk-boundary factorization. So we will first solve the latter axiom.

\paragraph{Solving the axiom of bulk-boundary factorization.}

We will write the general solution of this axiom in terms of some arbitrary
structure constants $B_{r,\eta}(j,\ell)$, and $\H$
conformal blocks built from known Liouville theory conformal
blocks. The relevant conformal blocks are most easily defined by
decomposing the boundary regime Ansatz (\ref{twoptcase}),
\begin{multline}
\la  V_{\a_1}(z_1)V_{\a_2}(z_2){}_{s_+}
B_\deg(y_1)_{s_-}B_\deg(y_2)_{s_+} \ra = 
\\
\sum_{\eta_1,\eta_2=\pm} 
\int d\beta\ B^L_{s_+}(\a_1,\beta-\tfrac{\eta_1}{2b})
C^L_{s_+}(Q-\beta+\tfrac{\eta_1}{2b},\tdeg\underset{s_-}{|} \beta) 
 B^L_{s_+}(\a_2,\beta-\tfrac{\eta_2}{2b})
C^L_{s_+}(Q-\beta+\tfrac{\eta_2}{2b},\tdeg\underset{s_-}{|} \beta) 
\\
\times \left(R^L_{s_-,s_+}(\beta)\right)^{-1}
{\cal G}_{\beta,\eta_1,\eta_2}(\alpha_1,\alpha_2|z_1,z_2,y_1,y_2)\ .
\label{vvbbdec}
\end{multline}
A basis of solutions of the Knizhnik-Zamolodchikov equations in the
  boundary regime is obtained by multiplying 
the conformal blocks ${\cal
  G}_{\beta,\eta_1,\eta_2}(\alpha_1,\alpha_2|z_1,z_2,y_1,y_2)$ with
  the prefactor (first line) of (\ref{twoptcase}), while assuming the relation
  (\ref{ajbl}) between $\H$ spins and Liouville momenta. We will still
  denote the resulting $\H$ conformal blocks as ${\cal
  G}_{\beta,\eta_1,\eta_2}(\alpha_1,\alpha_2|z_1,z_2,y_1,y_2)$, and
  represent them schematically as
\bea
{\cal
  G}_{\beta,\eta_1,\eta_2}(\alpha_1,\alpha_2|z_1,z_2,y_1,y_2) =
\psset{unit=.35cm}
\pspicture[](-5,-3)(5,3)
\blockhorl{\a_1}{\a_1}{\a_2}{\a_2}{\eta_1}{\eta_2}{\beta} 
\endpspicture
\ ,
\eea
where the wiggly lines \psset{unit=.22cm} \pspicture[](-.5,0)(.5,0) 
\thecoil(0,0)(0,2) \endpspicture\ denote degenerate fields of momentum
  $\deg$, and the discrete indices $\eta_i=\pm$ indicate the fusion
  channels $\beta-\tfrac{\eta_i}{2b}$ of these degenerate boundary
  fields $B_{\deg}$ with another 
 boundary field $B_\beta$.

The general solution of the bulk-boundary factorization axiom is obtained by
replacing the Liouville structure constants $B^L_{s_+}C^L_{s_+}$ in
eq. (\ref{vvbbdec}) with arbitrary quantities $B_{r,\eta}(j,\ell)$,
\bea
{\cal S}= \sum_{\eta_1,\eta_2=\pm} 
\int d\beta\ B_{r,\eta_1}(j_1,\ell)B_{r,\eta_2}
(j_2,\ell)\left(R^H_r(\ell)\right)^{-1} {\cal
  G}_{\beta,\eta_1,\eta_2}(\alpha_1,\alpha_2|z_1,z_2,y_1,y_2) \ .
\label{bbsol}
\eea
(Recall the relation (\ref{rhrl}) between the Liouville and $\H$ boundary
reflection coefficients.)
We have indeed chosen our basis of conformal blocks for its factorizing
behaviour in the boundary factorization limit,
\bea
\underset{z_1-\bz_1\rar 0}{\lim} \psset{unit=.35cm}
\pspicture[](-5,-3)(5,3)
\blockhorl{\a_1}{\a_1}{\a_2}{\a_2}{\eta_1}{\eta_2}{\beta} 
\endpspicture 
= 
|w-z_1|^{2\Delta_\beta}
\pspicture[](-5.3,-3)(2,3)
\rput[l]{110}(-2.5,0){\extline{r}{\a_1}}
\rput[l]{-110}(-2.5,0){\extline{r}{\a_1}}
\rput[l]{-90}(-1.5,2.2){\mycoil{t}{\scriptstyle \eta_1}}
\rput[l]{0}(-2.5,0){\intline{t}{}{2.5}}
\rput[l]{0}(0.3,0){$\beta$}
\endpspicture
\times
\pspicture[](-2,-3)(5,3)
\rput[l]{70}(2.5,0){\extline{l}{\a_2}}
\rput[l]{-70}(2.5,0){\extline{l}{\a_2}}
\rput[l]{-90}(1.5,2.2){\mycoil{t}{\scriptstyle \eta_2}}
\rput[l]{180}(2.5,0){\intline{t}{}{2.5}}
\rput[r]{0}(-0.3,0){$\beta$}
\endpspicture \ ,
\eea
where the two factors depend on $\beta,\eta_1,j_1,z_1,y_1,w$ and $\beta,
\eta_2,j_2, z_2,y_2,w$ respectively. Here $w$ is the position of the
intermediate channel field of momentum $\beta$ on the boundary of the disc.

The quantities $B_{r,\eta}(j,\ell)$ can be interpreted as the
bulk-boundary structure constants of the $\H$ model. For given values
of the bulk and boundary spins $j$ and $\ell$, there are {\it two} such
structure constants labelled by $\eta=\pm$. The reason for this fact,
and a detailed analysis of the $\H$ bulk-boundary two-point function,
are given in section \ref{secbulk}.

Therefore, thanks to the bulk-boundary factorization axiom, our task
is now reduced to determining the structure constants
$B_{r,\eta}(j,\ell)$, i.e. showing that they agree with the Liouville
structure constants in eq. (\ref{vvbbdec}). For this, we need the
continuity assumption.

\paragraph{Solving the continuity assumption.}

We recall that the continuity assumption determines the terms which involve the
$-\tfrac{1}{b}$ channel in the fusion product of the two boundary
degenerate fields (\ref{degbdyope}). 
In order to exploit this
assumption, it is therefore convenient to use a new basis of
conformal blocks
(where we omit the dependence on
$(\alpha_1,\alpha_2|z_1,z_2,y_1,y_2)$):
\bea
\psset{unit=.2cm} 
{\cal G}_{\beta,0} = 
\pspicture[](-10,-4)(10,4)
\psline(-6,0)(6,0)\psline(-8,-4)(-6,0)(-8,4)
\psline(8,-4)(6,0)(8,4) \psline(0,0)(0,3) \thecoil(-2.5,4)(0,3) 
\thecoil(0,3)(2.5,4) 
\rput[l](.5,1.5){$0$}
\rput[t](3,-.5){$\beta$} \rput[t](-3,-.5){$\beta$} 
\endpspicture
& , \hspace{1cm} & \psset{unit=.2cm} 
{\cal G}_{\beta,-\frac{1}{b},0} =
\pspicture[](-10,-4)(10,4)
\psline(-6,0)(6,0)\psline(-8,-4)(-6,0)(-8,4)
\psline(8,-4)(6,0)(8,4) \psline(0,0)(0,3) \thecoil(-2.5,4)(0,3) 
\thecoil(0,3)(2.5,4) 
\rput[l](.5,1.5){$-\frac{1}{b}$}
\rput[t](3,-.5){$\beta$} \rput[t](-3,-.5){$\beta$} 
\endpspicture\ ,
\nonumber
\\
\psset{unit=.2cm} 
{\cal G}_{\beta,-\frac{1}{b},+} =
\pspicture[](-10,-4)(10,4)
\psline(-6,0)(6,0)\psline(-8,-4)(-6,0)(-8,4)
\psline(8,-4)(6,0)(8,4) \psline(0,0)(0,3) \thecoil(-2.5,4)(0,3) 
\thecoil(0,3)(2.5,4) 
\rput[l](.5,1.5){$-\frac{1}{b}$}
\rput[t](3,-.5){$\beta-\frac{1}{2b}$} \rput[t](-3,-.5){$\beta+\frac{1}{2b}$} 
\endpspicture
& , \hspace{1cm} & \psset{unit=.2cm}  
{\cal G}_{\beta,-\frac{1}{b},-}  =
\pspicture[](-10,-4)(10,4)
\psline(-6,0)(6,0)\psline(-8,-4)(-6,0)(-8,4)
\psline(8,-4)(6,0)(8,4) \psline(0,0)(0,3) \thecoil(-2.5,4)(0,3) 
\thecoil(0,3)(2.5,4) 
\rput[l](.5,1.5){$-\frac{1}{b}$}
\rput[t](3,-.5){$\beta+\frac{1}{2b}$} \rput[t](-3,-.5){$\beta-\frac{1}{2b}$} 
\endpspicture \ .
\label{contblocks}
\eea
The relation to our previous basis of conformal blocks is
\bea
{\cal G}_{\beta,\eta, \eta} = F_{\eta,0}(\beta)\ {\cal
  G}_{\beta+\frac{\eta}{2b},0} +F_{\eta,-\frac{1}{b}}(\beta)\ {\cal
  G}_{\beta+\frac{\eta}{2b},-\frac{1}{b},0}  \sp {\cal G}_{\beta,\eta,
  -\eta} = {\cal G}_{\beta, -\frac{1}{b},-\eta}\ ,
\eea
for some Liouville fusing matrix elements $F_{\eta,0}(\beta), 
F_{\eta,-\frac{1}{b}}(\beta)$ which depend on $\beta$ but not on
$\a_1,\a_2$. (These fusing matrix elements are known explicitly, but
we do not need their precise form.) 

Let us rewrite the solution of the factorization axiom (\ref{bbsol})
in terms of such conformal blocks:
\begin{multline}
{\cal S} = 
 \int d\beta\ \left(R^H_r(\ell)\right)^{-1} 
\left(B_{r,+}(j_1,\ell)B_{r,-}(j_2,\ell){\cal
  G}_{\beta,-\frac{1}{b},-} +
B_{r,-}(j_1,\ell) B_{r,+}(j_2,\ell){\cal
  G}_{\beta,-\frac{1}{b},+} \right)  
\\
+ \sum_\eta \int d\beta\ \left(R^H_r(\ell)\right)^{-1}
B_{r,\eta}(j_1,\ell) B_{r,\eta}(j_2,\ell) 
\left( F_{\eta,0}(\beta)\ {\cal
  G}_{\beta+\frac{\eta}{2b},0} +F_{\eta,-\frac{1}{b}}(\beta)\ {\cal
  G}_{\beta+\frac{\eta}{2b},-\frac{1}{b},0}\right)\ .
\end{multline}
The continuity assumption determines the terms in ${\cal
  G}_{\beta,-\frac{1}{b},\pm}$, and therefore the values of
the products  $B_{r,+}(j_1,\ell)B_{r,-}(j_2,\ell)$ and $B_{r,-}(j_1,\ell)
  B_{r,+}(j_2,\ell)$. 
All our conformal blocks are indeed linearly independent, up to the
  identity of
 blocks labelled by momenta with identical conformal weights,
  for instance ${\cal G}_{\beta,-\frac{1}{b},+}
  = {\cal G}_{Q-\beta,-\frac{1}{b},-}$. One should also take into
  account corresponding identities among the structure constants,
  namely $B_{r,\eta}(j,\ell)=R_r^H(\ell)B_{r,-\eta}(j,-\ell-1)$.

The resulting values of $B_{r,+}(j_1,\ell)B_{r,-}(j_2,\ell)$ and $B_{r,-}(j_1,\ell)
  B_{r,+}(j_2,\ell)$ must be the ones appearing in the
  decomposition of our Ansatz (\ref{vvbbdec}),
because we already know the Ansatz to be a
  solution of the continuity constraints. This determines
  $B_{r,\pm}(j,\ell)$ up to a
 $j$-independent rescaling,
  $B_{r,\pm}(j,\ell) \rar f_r(\ell)^{\pm 1} B_{r,\pm}(j,\ell)$. A
  non-trivial 
  rescaling ($f_r(\ell)\neq \pm 1$) can however be excluded by
  exploiting the terms in ${\cal
  G}_{\beta+\frac{\eta}{2b},-\frac{1}{b},0}$, which are again determined by
  the continuity constraint. This shows that the Ansatz is the only
  solution to our axioms.

Therefore, our lack of control over the ${\cal
  G}_{\beta+\frac{\eta}{2b},0}$ terms has not prevented us from fully
  determining the bulk two-point function, thanks to the
 bulk-boundary factorization axiom. In the standard Cardy-Lewellen
  formalism, the bulk two-point function would be fully determined
  from the disc one-point and sphere three-point functions, and the
  bulk-boundary factorization axiom would then come as a
  consistency check on these quantities. In our case, this consistency
  check is weaker, because it can involve only the part of the axiom
  which we do not use for determining the bulk two-point function.

\paragraph{Generalization.}

This reasoning can be generalized to arbitrary $\H$ bulk correlators
on the disc. Indeed, the existence of a bulk regime where 
the $\H$ correlators are known (thanks to the bulk OPE) gives a
nontrivial content to the continuity assumption. Moreover, our
determination of the $\H$ bulk two-point function also yields the
knowledge of the $\H$ bulk-boundary two-point function. Therefore,
we can in principle apply the bulk-boundary OPE (\ref{bbope}) 
to arbitrary $\H$ bulk
correlators, which proves our main result (\ref{main}) for correlators of
bulk fields and boundary fields ${}_r\Psi^\ell_r$  which preserve the
boundary condition.  Boundary condition
changing operators ${}_r \Psi^\ell_{r'}$
are more challenging: we leave their case as a conjecture, which is supported
by our check of all the axioms, and the analysis of the boundary
two-point function in Section \ref{secbdy}.

\subsection{$\H$-Liouville relation in the $m$-basis} 

The $m$-basis relation may be useful for the study of the $\SLU$ coset
model, which is formally quite close to the $\H$-model in the $m$
basis. The relation is obtained by straightforward application of the
integral transforms (\ref{phijmbm}), (\ref{psiemeta}) to the $\mu$-basis result
(\ref{main}): 
\begin{multline}
\la \prod_{a=1}^n \Phi^{j_a}_{m_a,\bar{m}_a}(z_a) \prod_{b=1}^m
\Psi^{\ell_b}_{m_b, \eta_b}(w_b)\ra 
\\
\propto  \prod_{a=1}^n N^{j_a}_{m_a,\bar{m}_a} \prod_{b=1}^m
N^{\ell_b}_{m_b, \eta_b} \int_\R \frac{du}{|u|}\ \int
d^{2n'+m'}y\ \prod_{a=1}^n\left(\mu_a^{-m_a}\bmu_a^{-\bar{m}_a}\right)
\prod_{b=1}^m |\nu_b|^{-m_b}\sgn^{\eta_b}(\nu_b)
\\
\times |\Theta_{n,m}|^\frac{k}{2}  \la \prod_{a=1}^n V_{\a_a}
(z_a) \prod_{b=1}^m B_{\beta_b}(w_b)
\prod_{a'=1}^{n'} V_\deg (y_{a'}) 
\prod_{b'=1}^{m'} B_\deg (y_{b'}) \ra\ .
\label{mmain}
\end{multline}
The non-trivial content of the formula is the fact that the Jacobian
for Sklyanin's separation of variables (\ref{muofy}) (which gives
$\mu_a,\nu_b$ as a function of the positions $y_{a'},y_{b'}$ of the
degenerate fields) is 
$|u|^{-2} |\Theta_{n,m}| \prod_{a=1}^n|\mu_a|^2
\prod_{b=1}^m |\nu_b|$. The integral over $y$ should
be understood as spanning the whole range of complex or real values, and to
include the combinatorial factors due to the invariance of
$\mu_a,\nu_b$ under permutations of $y_{a'}$ or $y_{b'}$; for instance
in the case of the bulk two-point function $n=2,m=0$ we have $\int d^2
y \equiv \int_{\Im y_1>0} d^2 y_1+\frac12 \int_{\R^2} dy_1\ dy_2$. The
integral over $|u|$ can be performed explicitly knowing that
$\mu_a,\nu_b$ all have a factor $|u|$, the result is
$\delta(i\sum_{a=1}^n (m_a+\bar{m}_a) +i\sum_{b=1}^m m_b)$. 
(Recall that in the $\H$ model physical values of $m_a+\bar{m}_a$ and
$m_b$ are pure imaginary.)
The sum over
$\sgn u$ then affects the Liouville boundary parameters, which are
still given by eq. (\ref{srphi}) but kept implicit in our formula.
The normalization factors $N^{j_a}_{m_a,\bar{m}_a}, N^{\ell_b}_{m_b,
  \eta_b}$ are given in (\ref{njmbm}), (\ref{nlmeta}), and we do not
write the $j,\ell,m$-independent normalization factor.

A few cases are particularly simple. If $2n+m-2=0$ the $\H$-Liouville
relation does not involve Liouville degenerate fields. This happens
for the bulk one-point function ($n=1,m=0$) and the boundary two-point
function ($n=0,m=2$). If $2n+m-2=1$ the relation involves one boundary
degenerate field, and therefore no singularity can occur from the collision of
two degenerate fields. This happens for the bulk-boundary two-point
function ($n=1,m=1$) and the boundary three-point function ($n=0,m=3$).


\section{Boundary two-point function \label{secbdy}}


The boundary two-point function for open strings living on a single
$AdS_2$ brane is already known, eq. (\ref{ptpt}), and we reproduce it
here up to irrelevant factors:
\bea
\la \Psi^{\ell_1}(t_1|w_1) \Psi^{\ell_2}(t_2|w_2)\ra_r
=  \delta(\ell_1+\ell_2+1)
\delta(t_{12}) + \delta(\ell_1-\ell_2) \tilde{R}_{r}^{H}(\ell_1)
|t_{12}|^{2\ell_1} \ .
\label{slr2pt}
\eea
Up to a change of the reflection number $\tilde{R}_{r}^H(\ell_1)$,
this is actually the most general form of the two-point function which
is compatible with the $\SLR$ symmetry (\ref{slrsym}), if the
boundary fields
follow the standard $\SLR$ transformation rule (\ref{gplw}). 
And indeed, the equations in
\cite{pst01} which yielded that solution can also be used to derive
a boundary two-point function between different branes, which is of
the same form \cite{rib02b}.
The resulting reflection number $\tilde{R}^{H\ ?}_{r,r'}(\ell_1)$ however has branch
cuts as a function of the boundary spin $\ell$. While this is not an
inconsistency, this is certainly a strange feature. 

Our relation with
Liouville theory (\ref{main}) however predicts  
\begin{multline}
\la {}_r\Psi^{\ell_1}(t_1|w_1)_{r'} \Psi^{\ell_2}(t_2|w_2)_r \ra 
\\
= \delta(\ell_1+\ell_2+1)
\delta(t_{12}) + \delta(\ell_1-\ell_2) \tilde{R}_{r,r'}^{H}(\ell_1)
|t_{12}|^{2\ell_1} e^{-\frac12(k-2)(r-r')\sgn t_{12}} \ ,
\label{liou2pt}
\end{multline}
with the $t$-basis reflection number
\bea
\tilde{R}_{r,r'}^{H}(\ell) = \frac{\pi}{\G(2\ell+1)} \frac{ R^L_{ \frac{r}{2\pi b}
      +\frac{i}{4b} , \frac{r'}{2\pi b}-\frac{i}{4b}}
\left(\beta\right) }{\sin(\pi \ell
  -i\frac{r-r'}{2b^2})} = \frac{\pi}{\G(2\ell+1)} \frac{ R^L_{ \frac{r}{2\pi b}
      -\frac{i}{4b} , \frac{r'}{2\pi b}+\frac{i}{4b}}
\left(\beta \right) }{\sin(\pi \ell
  +i\frac{r-r'}{2b^2})} \ ,
\eea
with $\beta=b(\ell+1)+\frac{1}{2b}$. This reflection number is
meromorphic in $\ell$, with no hint of a branch cut. And the
factor $e^{-\frac12(k-2)(r-r')\sgn t_{12}}$ contradicts the $\SLR$ symmetry.

We will argue that (\ref{liou2pt}) is actually the correct
$\H$ boundary two-point function, and that the result of \cite{rib02b}
is incorrect because it relies on erroneous symmetry assumptions. We
will indeed show that the $\H$ boundary condition changing operators 
should not belong to representations of $\SLR$ but rather to
representations of the universal covering group $\USLR$. 

NB: In this section we omit the dependence of two-point function in
the worldsheet coordinates $w_1,w_2$. This dependence is always a
factor $|w_1-w_2|^{-2\Delta_{\ell_1}} $.

\subsection{$\USLR$ symmetry}

Let us investigate how the assumption of $\USLR$ symmetry would
constrain the boundary two-point function. To begin with, we study the
possible actions of that group on the boundary fields
${}_r\Psi^{\ell}(t|w)_{r'}$. 

Consider a timelike coordinate $T$ on $\USLR$ such that
$T(\id)=0$ and $T(-\id)=1$. (As a manifold, $\USLR$ is identical to
the Anti-de Sitter space $AdS_3$.) Then the set of $\USLR$ elements such that $0\leq
T<1$ can be identified with the group $\SLR/\{\id ,-\id\}$.  
We parametrize elements of $\USLR$ as 
$G=(g,[T])$ where $g$ is an element of the group $\SLR/\{\id
,-\id\}$, and $[T]$ is the integer part of $T$. 

The natural action of the group $\USLR$ on the parameter
$t$ is simply $(g,[T])\cdot t = g\cdot t$. It is however possible to
define an action of $\USLR$ on the $t$-basis fields $\Psi^\ell(t)$
which does not reduce to the ordinary $\SLR$ action $(g,[T])\cdot
 \Psi^\ell (t) = |ct-d|^{2\ell} \Psi^\ell(g\cdot t)$ as follows: for an $\USLR$ group
element $G=(g,[T])$ and a real number $t$ consider the number $N$ of
times $g\cdot t$ crosses $t=+\infty$ when $G$ continuously varies from
$G=\id_{\USLR}=(\id,0)$ to $G=(g,[T])$. Then for any fixed number $\kappa$
the following is an action of $\USLR$ on $t$-basis fields:
\bea
G\cdot \Psi^\ell(t) = 
(g,[T]) \cdot \Psi^\ell(t) = |ct-d|^{2\ell} 
e^{\kappa N(g,[T],t)} \Psi^\ell(g\cdot t)\ .
\label{gplt}
\eea
How would invariance under such $\USLR$ transformations constrain the
boundary two-point function?
Using $N(g,[T],t) =
[T]+\frac12+\frac12\sgn(t-d/c)$, we have
\begin{multline}
\la G\cdot {}_r \Psi^{\ell_1}(t_1|w_1)_{r'}\ G\cdot
    {}_{r'}\Psi^{\ell_2}(t_2|w_2)_r \ra
\\
= |ct_1-d|^{2\ell_1} |ct_2-d|^{2\ell_2} 
e^{\frac12 \kappa\left(\sgn[t_1-d/c]-\sgn[t_2-d/c]\right)} 
\la {}_r \Psi^{\ell_1}(g\cdot t_1|w_1)_{r'} \Psi^{\ell_2}(g\cdot
t_2|w_2)_r \ra  
\\
= |ct_1-d|^{2\ell_1} |ct_2-d|^{2\ell_2}
e^{\frac12 \kappa\left(\sgn t_{12}-\sgn [g\cdot t_1-g\cdot
    t_2]\right)} \la {}_r \Psi^{\ell_1}(g\cdot t_1|w_1)_{r'} \Psi^{\ell_2}(g\cdot
t_2|w_2)_r \ra \ .
\end{multline}
The requirement that this equals $ \la {}_r
\Psi^{\ell_1}(t_1|w_1)_{r'} \Psi^{\ell_2}(t_2|w_2)_r \ra  $ leads to 
\begin{multline}
\la {}_r\Psi^{\ell_1}(t_1|w_1)_{r'} \Psi^{\ell_2}(t_2|w_2)_r \ra 
\\
= \delta(\ell_1+\ell_2+1)
\delta(t_{12}) + \delta(\ell_1-\ell_2) \tilde{R}_{r,r'}^{H}(\ell_1)
|t_{12}|^{2\ell_1} e^{-\frac12\kappa\sgn t_{12}} \ ,
\label{sympred}
\end{multline}
for some $t$-basis reflection number
$\tilde{R}_{r,r'}^{H}(\ell_1)$. Therefore, the two-point function (\ref{liou2pt})
derived from the $\H$-Liouville relation is compatible
with $\USLR$ symmetry provided the boundary fields transform as
eq. (\ref{gplt}) with 
\bea
\kappa = (k-2)(r-r')\ .
\eea
We have thus found a nice geometrical interpretation for the two-point
function derived from the $\H$-Liouville relation. This is of course
not in itself evidence for the correctness of that relation. We will
look for such evidence in the comparison with N=2 Liouville theory,
and in the classical analysis of the $\H$ sigma model. 

\subsection{Comparison with N=2 Liouville theory}

An $\H$ mod $U(1)$ coset model can be obtained from the $\H$
model by gauging, and this coset model is known to be identical to 
the 2d black hole coset model $\SLU$ \cite{gaw91}.
It is also known that the N=2 supersymmetric version of the $\SLU$ coset
is related via mirror symmetry to $N=2$ Liouville
theory \cite{gkp99,kkk01,hk01}.
The boundary two-point function on maximally
symmetric D-branes in N=2 Liouville theory with central charge
$c=3+\frac{6}{k-2}$ is thus expected to be
related to the boundary two-point function on our $AdS_2$ branes in
the $\H$ model at level $k$. We
will not try to check this expectation in full detail, rather we will
focus on the non-trivial part of the expected relation, namely the
relation between the boundary reflection coefficients in the $\H$
model and N=2 Liouville theory.

The boundary reflection coefficient in N=2 Liouville theory
 was determined in \cite{hos04}. The D-branes
which should be compared to the $AdS_2$ branes in $\H$ are the
B-branes \cite{rib05b}. The relation between the parameters $r$ of our
$AdS_2$ branes and the parameters $J$ of the N=2 Liouville B-branes
can be deduced from the explicit formulas for the corresponding
one-point functions: $r=-\frac{i\pi}{k-2}(2J+1)$.
The boundary fields which span the
spectrum of open strings between such B-branes are called 
$ B^{\ell (s)}_m, \lambda B^{\ell (s)}_m, \bar{\lambda} B^{\ell
  (s)}_m, \lambda \bar{\lambda} B^{\ell (s)}_m$, where $\ell$ and $m$
correspond to the $\H$ boundary spin and $m$-basis momentum, $s$ is a
fermionic label which we will ignore because the $s$-dependence of the
N=2 Liouville boundary two-point function is trivial, and
$\lambda,\bar{\lambda}$ are boundary fermions such that
$\lambda\bar{\lambda} +\bar{\lambda}\lambda=1$. We will
compare the spectrum of open strings in $\H$ with the bosonic sector of the N=2
Liouville boundary spectrum; for each choice of $\ell,m$ this sector
is two-dimensional and spanned by $\bar{\lambda}\lambda B^{\ell (s)}_m, \lambda
\bar{\lambda} B^{\ell (s)}_m$. 

Let us write explicitly the reflection matrix for such N=2 Liouville
boundary fields \cite{hos04} (section 6.2 therein) 
with our notations and our own field normalizations chosen for later
convenience. (Changing field normalizations amounts to conjugating
the  matrix ${\cal M}$ with a diagonal matrix.)
\bea
\left(\begin{array}{c} \lambda \bar{\lambda} B^{\ell }_m
\\  \bar{\lambda}\lambda B^{\ell }_m \end{array}\right) &=& 
\frac{4}{\pi} \G(-\ell+m)\G(-\ell-m) \G(2\ell+1)
\tilde{R}^H_{r,r'}(\ell)\times {\cal M} \left(\begin{array}{c} \lambda
  \bar{\lambda} B^{-\ell-1 }_m 
\\  \bar{\lambda}\lambda B^{-\ell-1 }_m  \end{array}\right)\ ,
\label{lblrefl}
\eea
\bea
{\cal M}&=&
 \left( \begin{array}{cc}
\sum_\pm \pm e^{\mp\frac{r-r'}{2b^2}} \sin\pi(m\pm\ell)
  & e^{-i\pi m} e^{\frac{r-r'}{2b^2}} \sin 2\pi \ell \\
e^{i\pi m} e^{-\frac{r-r'}{2b^2}} \sin 2\pi \ell & \sum_\pm \mp
e^{\pm\frac{r-r'}{2b^2}} \sin\pi(m\pm\ell)  \end{array} 
\right) \ .
\nn
\eea
The $m$-basis boundary fields $\Psi^{\ell}_{m,\eta}$ of the $\H$ model
 were defined in (\ref{psiemeta}).
 Our $\H$ boundary two-point function (\ref{liou2pt}) has the following form in
the $m$-basis:
\begin{multline}
\la {}_r \Psi^{\ell_1}_{m_1,\eta_1}(w_1) _{r'} \Psi^{\ell_2}_{m_2,\eta_2}(w_2)_r
\ra =\delta(i(m_1+m_2))\times
\left[\delta(\ell_1+\ell_2+1) 2\pi \delta_{\eta_1\eta_2}
  \begin{array}{c} \\ \end{array} \right.
\\  + \delta(\ell_1-\ell_2) \frac{4}{\pi} 
  \G(-\ell_1+m_1) \G(-\ell_1-m_1) \cos
  \tfrac{\pi}{2}(\ell_1-m_1+\eta_1) \cos
  \tfrac{\pi}{2}(\ell_1+m_1+\eta_2)  
\\ \left. \times i^{\eta_1+\eta_2}
\G(2\ell_1+1) \tilde{R}^H_{r,r'}(\ell_1)  
  \left\{ (-1)^{\eta_1} \sin(\pi \ell+i\tfrac{r-r'}{2b^2})
  +(-1)^{\eta_2} \sin(\pi \ell-i\tfrac{r-r'}{2b^2}) \right\} \right] \ .
\label{mtwopt}
\end{multline}
 If we now assume the following identification between the N=2
 Liouville fields $\lambda\bar{\lambda}B^\ell_m,
 \bar{\lambda}\lambda B^\ell_m $ and the $\H$ model fields
 $\Psi^{\ell}_{m,\eta}$, 
which involves an implicit Wick rotation of the allowed values of $m$,
\bea
\lambda\bar{\lambda}B^\ell_m &\simeq& \Psi^\ell_{m,0}+\Psi^\ell_{m,1}
 =2\int_0^\infty dt\ t^{-\ell-1+m} \Psi^\ell(t) \ ,
\\
\bar{\lambda}\lambda B^\ell_m &\simeq & e^{i\pi
 m}(\Psi^\ell_{m,0}-\Psi^\ell_{m,1})=
  2e^{i\pi m} \int_{-\infty}^0 dt\
 |t|^{-\ell-1+m} \Psi^\ell(t)\ ,
\eea
then the $\H$ reflection matrix deduced from our $m$-basis
boundary two-point function (\ref{mtwopt}) 
agrees with the N=2 Liouville boundary reflection
matrix (\ref{lblrefl}).  

\subsection{Classical analysis }

We should be able to study
such a basic property of the theory of open strings in $\H$ as its
symmetry group without solving the full quantum theory. 
In the cases of closed strings and open strings which preserve 
boundary conditions, the
minisuperspace limit reduces our conformal field
theory to the quantum mechanics of a point particle in $\H$ and
$AdS_2$ respectively, and therefore
gives substantial insight into
the spectrum and symmetry properties. 
However, the theory of open strings stretched between two different
$AdS_2$ branes does not have such a minisuperspace limit, because such
open strings can not shrink to point particles. 
However, we will be able to gain some insight from 
analyzing their classical worldsheet dynamics.

In order to predict the symmetry group, we should derive the spectrum
of a timelike generator $R$ of the Lie algebra $s\ell_2(\R)$. (Such a
generator geometrically acts as a rotation of the $AdS_2$ branes.) Indeed,
such a generator must satisfy $\exp 2\pi iR=-\id$ if the symmetry
group is $\SLR$. On the other hand, no such relation exists in the
universal covering group $\USLR$. Nevertheless, the transformation law
(\ref{gplt})  of the boundary fields suggests that the value of $\exp
2\pi i R$ applied to such fields should be
$\exp 2\pi iR = e^{(k-2)(r-r')}$.
The operator $\exp 2\pi iR$ is indeed identified with the
$\USLR$ group element $G=(\id, 1)$, and for any real number $t$ we have
$N(\id,1,t)=1$. The spectrum of the quantum operator $R$ is therefore
expected to be 
\bea
Spec(R) = (k-2)\frac{r-r'}{2\pi i} + \Z\ .
\label{specr}
\eea
Of course, we do not expect the classical analysis to fully reproduce
this spectrum, and in particular not the $\Z$ quantization.
In order to show that the symmetry group is $\USLR$ and not $\SLR$, it
is enough to demonstrate that the spectrum is not purely real. We will
actually even find indications of an imaginary part proportional
to $r-r'$. 

In principle one can obtain the full set of classical solutions of the
$\H$ sigma-model, but it is not easy to extract predictions for the
spectrum of the rotation generator $R$.
This is due to the pure imaginary $B$-field in the theory
on worldsheets with Lorentzian signature which prevents classical
strings from evolving normally in time.
On the other hand, the model on Euclidean worldsheets has many
classical solutions, but it is not obvious how to relate the
spectrum of $R$ evaluated on classical solutions with the
quantum spectrum (\ref{specr}).
We will avoid  these subtleties by considering a 
classical solution which 
does not depend on the worldsheet time and therefore
makes sense for both signatures. 
Up to simple symmetry transformations, this is actually the unique
time-independent solution:
\bea
h=\exp  \Omega \left(r+ (r'-r)\frac{\sigma}{\pi}\right) \ ,
\eea
where $\Omega=\left(\begin{array}{cc} 0 & 1 \\ 1 &
  0\end{array}\right)$ and $\sigma $ is the space-like coordinate on
  the worldsheet. The complex coordinate on the upper half-plane
  worldsheet is $z=e^{\tau + i\sigma}$; our solution corresponds to
  inserting a boundary operator at $z=0$:
\bea
\psset{unit=.5cm}
\pspicture[](-5,-1.5)(5,2.5)
    \rput[l]{0}(-4,0){\bdy{8}}
    \rput[l]{0}(0,0){\genfield{l}{}}
    \rput[t]{0}(0,-.5){$\Psi$}
    \rput[t]{0}(2,-.5){$r$}
\rput[t]{0}(-2,-.5){$r'$}
\rput[t]{0}(0,2){$h(z,\bz)$}
\rput[l]{0}(4.3,0){$z=\bz$}
\endpspicture
\nonumber
\eea
Our solution is easily found to satisfy the following requirements:
\begin{enumerate}
\item Solving the bulk equations of motion. This is because $h$ can be
  factorized into holomorphic and antiholomorphic factors.

\item Solving the boundary conditions at $z=\bz$. 
In terms of the currents
\bea
J = k\p h h^{-1}  \sp \bar{J}=kh^{-1}\bp h 
= J^\dagger\ ,
\label{jhg}
\eea
these boundary conditions are of the type
\bea
J^\dagger \Omega^\dagger +\Omega J \underset{z=\bz}{=}0 \ .
\label{jooj}
\eea
This implies the vanishing of the derivative of
$\Tr \Omega h$ along the boundary, so that
\bea
\Tr \Omega h \underset{z=\bz}{=} \left\{ \begin{array}{ll} 2\sinh r & ,\ \Re z>0
\\ 2\sinh r' & ,\ \Re z<0 \end{array} \right.\ ,
\eea 
as required by the definition of the brane parameters $r,r'$ (\ref{troh}).

\item Corresponding to an affine primary field insertion at
  $z=0$. This means that the currents behave as
\bea
J(z) = \frac{k}{z}j_{0} + k\sum_{n=1}^\infty j_{-n} z^{n-1}\ .
\label{jzexp}
\eea
\end{enumerate}
We can now evaluate the values of the conserved momenta
associated to $s\ell(\R)$ transformations:
\bea
i\int_0^\pi e^\tau d\sigma\ \left(\Omega^{-1} J^\dagger \Omega + J\right) =
 kj_{0} = k\frac{r-r'}{2\pi i} \Omega \ .
\eea
The matrix $\frac12 \Omega$ satisfies $\exp 2\pi i (\frac12\Omega) = -\id $ and can
therefore be identified with the $R$ generator of the compact, timelike direction
of $s\ell (\R)$. The associated conserved charge of our classical
solution is
\bea
R^{cl} = k\frac{r-r'}{2\pi i}\ .
\eea
This agrees with the imaginary part of the spectrum of the quantum
operator $R$ (\ref{specr}), up a term which is subleading as $k\rar
\infty$. This is consistent with the classical analysis becoming
reliable only in the large $k$ limit, since $k$ appears as a factor in
the $\H$ action (\ref{H-act}) and therefore plays the r\^ole of the
inverse Planck constant.


\section{Bulk-boundary two-point function \label{secbulk}}


Like the boundary two-point function, the case of the bulk-boundary
two-point function will provide a nontrivial check of our 
 expression for $\H$ disc correlators in terms of
Liouville theory. We will indeed use a minisuperspace analysis to
independently predict the large level limit of the bulk-boundary
two-point function.

According to the formula (\ref{main}), the $\H$ bulk-boundary
two-point function is
\begin{multline}
  \la\Phi^j(\mu|z)\Psi^\ell(\nu|w)\ra_r
 \propto \delta(\mu+\bmu+\nu) |u| \left|
 \frac{(z-\bz)(z-w)(\bz-w)}{(w-y)(z-y)(\bz-y)}\right|^{\frac{k-2}{2}} 
\\ \times 
\la V_\a(z){}_{\frac{r}{2\pi b}+\frac{i}{4b}\sgn\nu}
 B_\beta(w)_{\frac{r}{2\pi b}-\frac{i}{4b}\sgn\nu} B_\deg (y) \ra\ ,
\label{ppvvb}
\end{multline}
where the Liouville momenta $\a,\beta$ are functions of $j,\ell$
(\ref{ajbl}), the position $y=-\frac{\mu \bz w +\bmu z w +\nu z
  \bz}{\mu z+\bmu\bz+\nu w}$ of the Liouville degenerate field is the zero
of the function $\varphi(t)$ (\ref{vphi}), we use $u=\mu z+\bmu
\bz+\nu w $ (\ref{udef}), and we omit the numerical factors. Here is a
picture of this $\H$-Liouville relation:
\bea
\psset{unit=.7cm}
\pspicture[](-3.5,-1)(3.5,3)
    \rput[l]{0}(-3,0){\bdy{6}}
    \rput[l]{0}(0,2){\genfield{l}{\Phi^j(\mu|z)}}
    \rput[l]{0}(0,0){\genfield{l}{\Psi^\ell(\nu|w)}}
    \rput[t]{0}(-2,-.5){$r$}
\rput[t]{0}(2,-.5){$r$}
      \endpspicture
\simeq  
\psset{unit=.7cm}
\pspicture[](-8,-1.5)(5,3)
    \rput[l]{0}(-7.5,0){\bdy{12}}
    \rput[l]{0}(0,2){\genfield{l}{V_\a(z)}}
    \rput[l]{0}(0,0){\genfield{l}{B_\beta(w)}}
    \rput[l]{90}(-3,0){\degfield{r}{B_\deg(y)}}
\rput[t]{0}(-5.5,-.5){$\frac{r}{2\pi b} - \frac{i}{4b} \sgn\nu$}
\rput[t]{0}(-1.5,-.5){$\frac{r}{2\pi b} + \frac{i}{4b} \sgn\nu$}
\rput[t]{0}(2.5,-.5){$\frac{r}{2\pi b} - \frac{i}{4b} \sgn\nu$}
      \endpspicture
\nonumber
\eea
The Liouville boundary parameter is therefore controlled by $\nu$,
which we spell out explicitly in terms of the separated variables
$(u,y)$ thanks to eq. (\ref{muofy}):
\bea
\nu = u\frac{w-y}{|w-z|^2}\ .
\eea

\subsection{$\SLR$ symmetry}

We first check that our formula for the bulk-boundary two-point
function obeys the $\SLR$ symmetry requirement (\ref{slrsym}). 
The general solution to this requirement is
\begin{multline}
  \la\Phi^j(x|z)\Psi^\ell(t|w)\ra_r =
 |z-w|^{-2\Delta_\ell}|z-\bar z|^{-2\Delta_j+\Delta_\ell}
\\
\times 
|x+it|^{2\ell}|x+\bar x|^{2j-\ell} \frac{\G(-2j-\ell-1)}{2\pi}
 \sum_{\pm} B^H_{r,\pm}(j,\ell)\ e^{\pm i\frac{\pi}{2}(2j+\ell+1)
 \sgn\Re x}\ .
\label{pjpl} 
\end{multline}
Like in the case of the bulk one-point function (which is obtained 
for $\ell=0$), the $\SLR$ symmetry allows an arbitrary dependence
on $\sgn \Re x$. Here we choose $e^{\pm i\frac{\pi}{2}(2j+\ell+1)
 \sgn\Re x}$ as a basis of functions of $\sgn\Re x$, and we introduce
the {\it two $\H$ bulk-boundary structure constants}
$B^H_{r,\pm}(j,\ell)$. The factor $\frac{\G(-2j-\ell-1)}{2\pi}$ is
chosen for later convenience. 

We now transform this bulk-boundary two-point function into the
$\mu$-basis (defined by equations (\ref{pmupx}), (\ref{pnupt}))
for the purpose of the comparison with the formula predicted by
our $\H$-Liouville relation.
The Fourier integral over $(x,t)$ can be performed 
by making the change of variables $x=x'-it$ and then parametrizing
$x'\in \C$ in terms of real variables $\sigma,\tau$ such that $x'=
\sigma(i\tau-\bmu) $. (Then the integral over $\sigma$ is of the type
(\ref{itnu}).)
\begin{multline}
 \la\Phi^j(\mu|z)\Psi^\ell(\nu|w)\ra_r =
 |z-\bar z|^{\Delta_\ell-2\Delta_j}|z-w|^{-2\Delta_\ell}
\\
 \times \delta(\mu+\bar\mu+\nu)\ 
 |\mu|^{2j+2}|\nu|^{-\ell}
 \int_{-\infty}^\infty d\tau\ |\tau|^{-2j-\ell-2}|\tau-i\mu|^{2\ell}
 B^H_{r,\sgn\tau}(j,\ell)\ .
\label{Irjlmunu}
\end{multline}
The remaining integral over $\tau$ converges provided $\mu$ is not
pure imaginary. It can be performed using the integral formula
(\ref{ixab}) which yields:
\begin{multline}
 \la\Phi^j(\mu|z)\Psi^\ell(\nu|w)\ra_r =
 |z-\bar z|^{\Delta_\ell-2\Delta_j}|z-w|^{-2\Delta_\ell}
\\
 \times \delta(\mu+\bar\mu+\nu)\ 
 |\mu|^{\ell+1}|\nu|^{-\ell}
 \frac{\G(-2j-1-\ell)\G(2j+1-\ell)}{\G(-2\ell)}  \sum_{\pm}
 B^H_{r,\pm}(j,\ell) F^\pm_{j,\ell}(\mu)\ ,
\label{sbf}
\end{multline}
where we define
\bea
F_{j,\ell}^\pm(\mu) \equiv
\left(\frac12\pm\frac12\frac{\Im\mu}{|\mu|} \right)^{\ell+\frac12}
F\left(2j+\frac32, -2j-\frac12,\frac12-\ell;
\frac12\mp\frac12\frac{\Im\mu}{|\mu|}\right) \ .
\label{fjlpm}
\eea

The Liouville correlator in (\ref{ppvvb}) can be decomposed into Liouville
structure constants, and conformal blocks which capture all the
dependence on the worldsheet coordinates $z,w,y$.
The properties of the relevant blocks have been studied in \cite{hos01},
and they are proportional to the functions
$F_{j,\ell}^\pm(\mu)$ in (\ref{fjlpm}).
\footnote{According to \cite{hos01}, the relevant Liouville blocks are
  indeed powers of $|z-w|,|z-y|,|w-y|$, times
hypergeometric functions of the type 
$F(b^{-1}(2\a+\beta-Q),b^{-1}(\beta-\tfrac{1}{2b}),2b^{-1}(\beta-\tfrac{1}{2b})
;\tilde{z})$ with $\tilde{z}=
\frac{(z-\bz)(y-w)}{(z-w)(y-\bz)}=1+\frac{\mu}{\bmu}$. 
The relation with $F_{j,\ell}^\pm(\mu)$ is established thanks to the
quadratic transformation (\ref{fquad}).}
What is however not obvious, but
necessary for the $\SLR$ symmetry, is that the coefficients of this
decomposition are completely independent of $\mu,\nu$, in spite of the
$\sgn \nu$ dependence of the Liouville boundary parameter.

In order to write the decomposition explicitly,
let us consider the Liouville 
factorization limit $y\rar w$ when the degenerate
boundary field $B_\deg$ collides with $B_\beta$. This limit
  corresponds to $\nu=0$, and therefore $\mu$ pure imaginary 
(using $\mu+\bmu+\nu=0$). The behaviour of our conformal blocks
  $F^\pm_{j,\ell}(\mu)$ in this limit is actually determined by 
$\underset{\nu\rar 0}{\lim}\ \sgn \Im \mu = -\sgn u$. Namely, $F^+$ is
  regular if $\sgn u =-$ and $F^-$ is regular if $\sgn u =+$. The
  Liouville correlator in (\ref{ppvvb}) is then decomposed into
  regular blocks, and structure constants where the boundary
  parameters can be determined from the identity $\sgn\nu =\sgn u\
  \sgn(w-y)$. 
We thus find the following decomposition, where $\e=\sgn u$:
\begin{multline}
\la\Phi^j(\mu|z)\Psi^\ell(\nu|w)\ra_r =
 |z-\bar z|^{\Delta_\ell-2\Delta_j}|z-w|^{-2\Delta_\ell}
 \delta(\mu+\bar\mu+\nu)
\\
\times \left[
C^L_{s_\e}(\beta \underset{s_{-\e}}{|}\tdeg , Q-\beta-\tfrac{1}{2b})
 B_{s_\e}^L(\a,\beta+\tfrac{1}{2b}) 
 |\mu|^{\ell+1}|\nu|^{-\ell} F^{-\e}_{j,\ell}(\mu)
 \right.\\ \left.
+C^L_{s_\e}(\beta \underset{s_{-\e}}{|}\tdeg , Q-\beta+\tfrac{1}{2b})
 B_{s_\e}^L(\a,\beta-\tfrac{1}{2b})
 |\mu|^{-\ell}|\nu|^{\ell+1} F^{-\e}_{j,-\ell-1}(\mu)  \right] \ ,
\label{twodec}
\end{multline}
where $s_\pm=\frac{r}{2\pi
  b}\mp\frac{i}{4b}$, and 
the Liouville bulk-boundary structure constant $B_s^L(\a,\beta)$
(\ref{vbs}) and degenerate boundary three-point structure constant
\mbox{$C^L_{s}(\beta \underset{s'}{|}\tdeg , Q-\beta\pm\tfrac{1}{2b})$}
(\ref{bbbss}) are explicitly known.

The Liouville correlator in (\ref{ppvvb}) is known to have an
alternative decomposition \cite{hos01}, which leads to equation
(\ref{twodec}) being also valid for $\e=-\sgn u$. (The equality of these
decompositions can be exploited in order to derive
a $\frac{1}{2b}$-shift relation for the Liouville bulk-boundary 
structure constant.) We will now use these two decompositions
$\e=\pm$, while rewriting
the functions $F^{-\epsilon}_{j,-\ell-1}(\mu)$ in terms of
$F^\pm_{j,\ell}(\mu)$ with the help of
\bea
\left|\frac{\nu}{4\mu}\right|^{2\ell+1} \frac{\G(-\ell+\tfrac12)
  \G(-\ell-\tfrac12) }{\G(-\ell-2j-1) \G(-\ell+2j+1)} F^\e_{j,-\ell-1}
= F^{-\e}_{j,\ell} +\frac{\cos \pi 2j}{\cos \pi \ell} F^\e_{j,\ell}\ .
\eea
Not forgetting
$C^L_{s}(\beta \underset{s'}{|}\tdeg , Q-\beta+\tfrac{1}{2b})=1$, we
find that the $\H$ bulk-boundary two-point function deduced from the
$\H$-Liouville relation is indeed of the form (\ref{sbf}) dictated by $\SLR$
symmetry, with structure constants
\begin{equation}
 B^H_{r\pm}(j,\ell) =
 \frac{2^{4\ell+2}\Gamma(-2\ell)}{\Gamma(-\ell-\frac12)\Gamma(-\ell+\frac12)}
 B^L_{\frac{r}{2\pi b}\mp\frac{i}{4b}}(\alpha,\beta-\tfrac{1}{2b})\ .
\label{bhbl}
\end{equation}
(We omit numerical factors.)

\subsection{Minisuperspace analysis}

Let us compute the minisuperspace limit $k\rar \infty$ 
of our $\H$ bulk-boundary two-point function. Thanks to $\SLR$
symmetry, this reduces to computing the $k\rar \infty$ limit of the
structure constants $B^H_{r,\pm}(j,\ell)$ computed in \cite{hos01} and
reproduced in the Appendix, eq. (\ref{bhbl}). 
Using the explicit formula for the Liouville bulk-boundary structure
constants $B^L$ (\ref{vbs}), we compute
their semi-classical limit $b^2=\frac{1}{k-2}\rar 0$ (assuming
$r,j,\ell$ stay fixed): 
\begin{eqnarray}
\lefteqn{
 \lim_{b\to 0}B^L_{\frac{r}{2\pi b}\mp \frac{i}{4b}}(\alpha,\beta-\tfrac{1}{2b})
 ~=~ 4\pi(-\mu_L\pi b^{-2}e^{2r})^{-1-j-\frac\ell2}
} \nn\\ &&
  \int_{i\R}\frac{dp}{2\pi i} e^{(-2r\pm i\pi)p}
  \frac{\Gamma( p+\ell+2j+2)\Gamma(p+\ell+1)\Gamma(-p-2j-1)\Gamma(-p)}
       {\Gamma(\ell+1)\Gamma(-2j-1)}
 \nn\\
 &=& 4\pi(\mu\pi b^{-2})^{-1-j-\frac\ell2}
 \nn\\ && \times
  \left\{
   e^{\mp\frac{i\pi}{2}(2j-\ell+2)}
    \frac{\Gamma(-2j+\ell)\Gamma(2j+2)}{\Gamma(-2j)}
    e^{(2j-\ell)r}
    F(\ell+1,-2j+\ell,-2j;-e^{-2r})
 \right. \nn\\ && \left.
 +~e^{\pm\frac{i\pi}{2}(2j+\ell+2)}
    \Gamma(2j+\ell+2)
    e^{-(2j+\ell+2)r}
    F(\ell+1,2j+\ell+2,2j+2;-e^{-2r})
  \right\}\ ,
\label{limpred}
\end{eqnarray}
where we make use of the asymptotic behaviour of the special functions
(\ref{limgb})-(\ref{limsbr})
and of the auxiliary formula
\bea
e^{(-2r\pm i\pi)p} = e^{-2r} \frac{e^{\mp i\pi
    \ell}\sin\pi(2j+p)-e^{\mp i\pi 2j}\sin \pi(\ell+p)}{\sin \pi
  (2j-\ell)}\ .
\eea

Let us now predict  
 the minisuperspace limit of the bulk-boundary structure constants
$B^H_{r,\pm}(j,\ell)$ 
by an independent calculation. In the minisuperspace
 limit, the $\H$ model path-integral reduces 
to the integral over worldsheet-independent elements $h$ of $\H$:
\bea
 \la\Phi^j(x|z)\Psi^\ell(t|w)\ra_r^{\rm mini}\equiv
\int_{\H} dh\
 \Phi^j(x|h)\Psi^\ell(t|h)
 \delta({\rm Tr}\left[h\Omega\right]-2\sinh r)\ .
\eea
Using the explicit formulas for the $\H$ element $h$ (\ref{hmat}) and
the classical fields $\Phi^j(x|h)$ (\ref{phijcl}) and $\Psi^\ell(t|h)$
(\ref{psilcl}),
the minisuperspace computation is performed as follows:
\begin{eqnarray}
\lefteqn{
 \la\Phi^j(x|z)\Psi^\ell(t|w)\ra_r^{\rm mini} 
} \nn\\
&=& -\frac{2j+1}{\pi}
 \int d^2\gamma\ d\phi\ e^{2\phi}\delta(e^\phi(\gamma+\bar{\gamma})-2\sinh r)
 (|x-\gamma|^2e^\phi+e^{-\phi})^{2j}
 (|it+\gamma|^2e^\phi+e^{-\phi})^{\ell}
\nn\\ &=&
 -\frac{2j+1}{2\pi}
 \int \frac{d^2\gamma du}{u^{2j+\ell+1}} \delta(\Re\gamma-\sinh r)
 (|u(x+it)-\gamma|^2+1)^{2j}
 (|\gamma|^2+1)^{\ell}
\nn\\ &=&
 |x+\bar x|^{2j-\ell}|x+it|^{2\ell}
 \nn\\ && \times
 \frac{\Gamma(-4j-1)(2\cosh r)^{2j+\ell+1}}{\Gamma(-2j)\Gamma(-2j-1)}
 \int_0^\infty \frac
 {du}{u^{2j+\ell+2}}(u^2-2u\tanh r\ \sgn \Re x+1)^{2j+\frac12},
\end{eqnarray}
where $\gamma$ was shifted $\gamma\rar \gamma-it$ and rescaled
$\gamma\rar e^{-\phi} \gamma$, 
we introduced $u=e^\phi$, 
and we reached the last expression by the rescaling
$u\rar \frac{(1+|\gamma|^2)\Re x}{|x+it|^2\cosh r} u$
which allowed the integral over $\gamma$ to be performed.

The remaining integral over $u$ can be performed with the help
of the formula (\ref{ixab}). The minisuperspace bulk-boundary
two-point function is then found to be of the form dictated by the
spacetime $\SLR$ symmetry (\ref{pjpl}), with the
structure constants:
\begin{multline}
 B^{H,{\rm mini}}_{r,\pm}(j,\ell) = (2\cosh r)^{2\ell+1} 
 \\ \times \left\{
 e^{\mp\frac{i\pi}{2}(2j-\ell+2)}
  \frac{\Gamma(-2j+\ell)\Gamma(2j+2)}{\Gamma(-2j)}
  e^{(2j-\ell)r}
  F(\ell+1,-2j+\ell,-2j;-e^{-2r})
 \right.
\\ \left. 
 +e^{\pm\frac{i\pi}{2}(2j+\ell+2)}
  \Gamma(2j+\ell+2)
  e^{-(2j+\ell+2)r}
  F(\ell+1,2j+\ell+2,2j+2;-e^{-2r})
\right\}\ .
\label{minipred}
\end{multline}
Up to a renormalization of the fields, this agrees with the prediction
(\ref{limpred}) from our $\H$-Liouville relation.

\appendix


\section{Appendix \label{secap} }


\subsection{Special functions}

The function $\gamma(x)$ is built from Euler's Gamma function:
\bea
\gamma(x)=\frac{\Gamma(x)}{\Gamma(1-x)}\,.
\eea
We use the special functions $\G_b$, $\up$ and $S_b$ which usually appear in
the study of Liouville theory at parameter $b>0$ and background charge
$Q=b+b^{-1}$. We use the same conventions as \cite{pon03}, where some
more details can be found. The following definitions are valid for
$0<\Re x<Q$:
\bea
\text{log}\G_b(x)&=&\int_{0}^{\infty}\frac{dt}{t}\left\lbrack
\frac{e^{-xt}-e^{-Qt/2}}{(1-e^{-bt})(1-e^{-t/b})}-
\frac{(Q/2-x)^{2}}{2}e^{-t}-\frac{Q/2-x}{t}\right\rbrack \, ,
\label{gammab}
\\
\text{log}\up&=&\int_{0}^{\infty}\frac{dt}{t}\left\lbrack
\left(\frac{Q}{2}-x\right)^{2}e^{-t}-
\frac{\text{sinh}^{2}(\frac{Q}{2}-x)\frac{t}{2}}{\text{sinh}\frac{bt}{2}
\text{sinh}\frac{t}{2b}}\right\rbrack \, ,
\label{upb}
\\
\text{log}S_b&=&\int_{0}^{\infty}\frac{dt}{t}\left\lbrack
\frac{\text{sinh}(\frac{Q}{2}-x)t}
{2\text{sinh}(\frac{bt}{2})\text{sinh}(\frac{t}{2b})}-
\frac{(Q-2x)}{t}\right\rbrack\ .
\label{sb}
\eea
These functions can be extended to a meromorphic function on the complex
plane thanks to the shift equations
\bea
\G_b(x+b)= \frac{\sqrt{2\pi}b^{bx-\frac{1}{2}}}{\Gamma(bx)}\G_b(x)  &\sp&
\G_b(x+1/b)=
\frac{\sqrt{2\pi}b^{-\frac{x}{b}+\frac{1}{2}}}{\Gamma(x/b)}\G_b(x)
\\
\up(x+b)=\frac{\Gamma(bx)}{\Gamma(1-bx)}b^{1-2bx}\up(x) &\sp &
\up(x+1/b)=\frac{\Gamma(x/b)}{\Gamma(1-x/b)}b^{2x/b-1}\up(x)
\\
S_b(x+b) = 2\text{sin}(\pi bx)S_b(x) & \sp & S_b(x+1/b) = 2\text{sin}(\pi
x/b)S_b(x)
\eea
The three special functions are related: $S_b(x)=
\frac{\G_b(x)}{\G_b(Q-x)} $ and $ \up(x) = \frac{1}{\G_b(x)\G_b(Q-x)}$.

Using the integral representations for the special functions, one can
study their behaviour for $b\rar 0$ while keeping the quantity $x$
fixed:
\bea
\Gamma_b(bx)\rar (2\pi b^3)^{\frac12 (x-\frac12)} \Gamma(x)  &\sp&
\Gamma_b(Q-bx)\rar \left(\frac{b}{2\pi}\right)^{\frac12 (x-\frac12)} \ ,
\label{limgb}
\\
\up(bx)&\rar& \frac{1}{b^{x-\frac12} \Gamma(x)} \ ,
\label{limupb}
\\
 S_b(bx)\rar (2\pi
b^2)^{x-\frac12} \Gamma(x) &\sp& S_b(\frac{1}{2b}+bx)\rar
2^{x-\frac12}\ ,
\label{limsb}
\\
\prod_\pm S_b(\frac{1}{2b}+bx\pm i \frac{r}{\pi b})
&\rar& \left(\frac{\cosh r}{\sqrt{2}}\right)^{1-2x} \ .
\label{limsbr}
\eea

\subsection{Useful formulas}

The following formula \cite{gr65} is used in Section \ref{secbulk}.
\begin{multline}
\int_0^\infty dx x^\alpha(1\pm 2x\tanh r+x^2)^\beta
 = \frac{\Gamma(\alpha+1)\Gamma(-2\beta-\alpha-1)}{\Gamma(-2\beta)}
\\ \times   
 (2\cosh r)^{-\beta-\frac12}
    e^{\pm r(\beta+\frac12)}
    F\left(-\alpha-\beta-\frac12,\alpha+\beta+\frac32,
       \frac12-\beta;\frac{1}{e^{\pm 2r}+1}\right)\ .
\label{ixab}
\end{multline}
The following formulas are useful for transforming some correlators in
the $(x,t)$ basis into correlators in the $(\mu,\nu )$ basis.
\bea
\int_\C d^2x\ e^{i\Im \mu x} |x|^{2\a} &=& \pi \gamma(\a+1)
|\mu|^{-2\a-2}\ , 
\label{ixmu}
\\
\int_\R dt\ f(\sgn t)|t|^\a e^{-it\nu}
&=& |\nu|^{-\a-1}\G(\a+1)\left[ f(\sgn \nu)e^{-i\frac{\pi}{2}(\a+1)}
    +f(-\sgn\nu) e^{i\frac{\pi}{2}(\a+1)} \right]\ .
\label{itnu}
\eea
The conformal blocks which are relevant for Section \ref{secbulk}
involve hypergeometric functions which can undergo a quadratic
transformation:
\begin{multline}
F(a,b,2b;z)=\left(\tfrac12+\tfrac14\tfrac{2-z}{\sqrt{1-z}}\right)^{\frac12-b}
(1-z)^{-\frac{a}{2}}
\\
\times
F\left(b-a+\tfrac12,a-b+\tfrac12,b+\tfrac12;
\tfrac12-\tfrac14\tfrac{2-z}{\sqrt{1-z}}\right)\ .
\label{fquad}
\end{multline}

\subsection{Some Liouville theory formulas}

We mostly follow conventions of \cite{pon03}. We consider Liouville
theory with parameter $b>0$, background charge $Q=b+b^{-1}$, 
central charge $c=1+6Q^2$, and interaction strength $\mu_L$.

One-point function on a disc:
\begin{multline}
\la V_\a(z)\ra_s=|z-\bar{z}|^{-2\Delta_\a} (\pi \mu_L
\gamma(b^2))^{\frac{Q-2\a}{2b}}
\\ \times
\frac{\Gamma(1-b(Q-2\a))\Gamma
(1-b^{-1}(Q-2\a))}{-\pi 2^\frac14 (2\a-Q)}\cosh 2\pi s (Q-2\a)\ .
\label{ld11}
\end{multline}
Boundary reflection coefficient and two-point function:
\begin{multline}
\la {}_s B_{\beta_1}(w_1)_{s'}B_{\beta_2}(w_2)_s\ra
=|w-\bar{w}|^{-2\Delta_{\beta_1}} \left[
  \delta(\beta_1+\beta_2-Q)+R^L_{s,s'}(\beta_1)\delta(\beta_1-\beta_2)\right]\ ,
\\
R^L_{s,s'}(\beta)=\left[\pi
  \mu_L\gamma(b^2)b^{2-2b^2}\right]^{\frac{Q-2\beta}{2b}}
\frac{\Gamma_b(2\beta-Q)}{\Gamma_b(Q-2\beta)}\prod_{\pm \pm} S_b\left(Q-\beta +
i(\pm s\pm s')\right) \ .
\label{rlssp}
\end{multline}
Bulk-boundary two-point function \cite{hos01}:
\begin{multline}
\la
V_\a(z)B_\beta(w)\ra_s=
|z-\bz|^{\Delta_\beta-2\Delta_\a}|z-w|^{-2\Delta_\beta}
B^L_s(\a,\beta)\ ,
\\
B^L_s(\a,\beta)=i2^{-\frac14}
\left[ \pi \mu_L \gamma(b^2)b^{2-2b^2}\right]^{\frac{Q-2\a-\beta}{2b}}
\frac{ \G_b^3(Q-\beta) \G_b(2Q-2\a-\beta) \G_b(2\a-\beta)
}{\G_b(Q)\G_b(\beta)\G_b(Q-2\beta) \G_b(2\a)\G_b(Q-2\a)}
\\
\times
\int_{-i\infty}^{i\infty}dp\ e^{-4\pi sp} \prod_\pm
\frac{S_b(\a+\frac{\beta-Q}{2}\pm p)}{S_b(\a-\frac{\beta-Q}{2}\pm p)}
\ .
\label{vbs}
\end{multline}
From this, one can deduce the bulk-boundary OPE of a bulk degenerate
field $V_{\deg}$ which is relevant for our continuity assumption
(\ref{degbbope}). There is a subtlety: due to the pole
structure of $B^L_s(\a,\beta)$, the degenerate 
limit of the bulk-boundary OPE yields
\bea
B^L_s(\tdeg,Q)=\underset{\a\rar \deg}{\lim}\
\underset{\beta=b-2\a}{\Res}\ B^L_s(\a,\beta)\ ,
\label{limres}
\eea
instead of the incorrect formula $B^L_s(\tdeg,Q)\overset{?}{=}
\underset{\a\rar \deg}{\lim}\ \underset{\beta=Q}{\Res}\
B^L_s(\a,\beta)$ which one might naively have expected. The correct
result is 
\bea
B^L_s(\tdeg,Q)=2b^{-2} \left[\pi
  \mu_L\gamma(b^2)\right]^{\frac{1}{2b^2}}
\frac{\G(-1-2b^{-2})}{\G(-b^{-2})} \cosh 2\pi b^{-1}s\ .
\label{vbbbb}
\eea
Operator product expansion of a degenerate boundary field:
\begin{multline}
_{s_+}B_\beta(w)_{s_-}B_\deg (y)_{s_+} \sim 
|w-y|^{b^{-1}(Q-\beta)} 
C^L_{s_+}(\beta \underset{s_-}{|}\tdeg , Q-\beta-\tfrac{1}{2b}) 
_{s_+}B_{\beta+\tfrac{1}{2b}}(w)_{s_+} 
\\
+|w-y|^{b^{-1}\beta}
C^L_{s_+}(\beta \underset{s_-}{|}\tdeg , Q-\beta+\tfrac{1}{2b})
_{s_+}B_{\beta-\tfrac{1}{2b}}(w)_{s_+} \ ,
\label{bbcbcb}
\end{multline}
with the degenerate boundary structure constants
\bee
C^L_{s_+}(\beta \underset{s_-}{|}\tdeg , Q-\beta+\tfrac{1}{2b}) &=&1 
\\
C^L_{s_+}(\beta \underset{s_-}{|}\tdeg , Q-\beta-\tfrac{1}{2b}) &=&
R^L_{s_-,s_+}(\beta) R^L_{s_+,s_+}(Q-\beta-\tfrac{1}{2b}) 
=\frac{2b^{-2}}{\pi}
\left[\pi \mu_L\gamma(b^{-1})\right]^\frac{1}{2b^2} \times
\eee
\vspace{-9mm}
\bea
\times  \G(1-2b^{-1}\beta)\G(2b^{-1}\beta-b^{-1}Q) 
\cos\pi(b^{-1}\beta-\tfrac{b^{-1} Q}{2})  \cos
\pi(b^{-1}\beta-\tfrac{b^{-1}Q}{2} \mp 2ib^{-1}s_+) \ ,
\label{bbbss}
\eea
where $s_+=s_- \pm \frac{i}{2b}$. The particular case of the OPE of two
degenerate boundary fields
$\beta=\tdeg$ is relevant for our continuity assumption
(\ref{degbdyope}). 



\acknowledgments{We are grateful to Volker Schomerus and Joerg
  Teschner for useful discussions, collaboration at some stages of
  this work, and comments on the draft of the article. 
  We thank each other's institutes for
  hospitality. S. R. is supported by a Humboldt stipendium of the
  Alexander von Humboldt Stiftung.  
}


\providecommand{\href}[2]{#2}\begingroup\raggedright\endgroup


\end{document}